\documentclass[12pt,english]{article}
\pdfoutput=1

\usepackage[T1]{fontenc}
\usepackage[utf8]{inputenc}
\usepackage{lmodern, amsmath,amssymb,extarrows, graphicx, pifont, adjustbox, bm, xcolor}
\usepackage{amsfonts}
\usepackage{geometry}
\usepackage{comment,mdframed,ulem}
\usepackage{mathtools}
\usepackage{float}
\usepackage{slashed}
\usepackage{cite}
\usepackage{ragged2e}
\usepackage{etoolbox}
\usepackage{array}
\usepackage{enumitem}
\usepackage{cancel}
\usepackage{braket}

\usepackage[unicode=true,pdfusetitle,
 bookmarks=true,bookmarksnumbered=false,bookmarksopen=false,
 breaklinks=false,pdfborder={0 0 1},backref=false,colorlinks=true]
 {hyperref}

\hypersetup{pdfauthor={Allic Sivaramakrishnan},
citecolor=black,linkcolor=black,urlcolor=black}

\usepackage{bm}  
\usepackage{braket}
\usepackage{slashed}
\usepackage{comment}
\usepackage{graphicx}
\usepackage{xcolor}
\usepackage{dcolumn} 
\usepackage{tensor} 
\usepackage{enumitem}
\usepackage{authblk}
\usepackage{siunitx}
\usepackage{mathtools}
\usepackage{dsfont}
\usepackage{ulem}
\usepackage{tikz}
\usetikzlibrary{arrows.meta, calc}
\usetikzlibrary{decorations.pathmorphing}
\usepackage{amsmath,amsthm,amsfonts,amscd,amssymb}

\newcommand\scri{\mathscr{I}}

\usepackage{dsdshorthand}

\newcommand{\mb}{\mathbf}

\begin{document}

\interfootnotelinepenalty=10000
\baselineskip=18pt

\vspace{2cm}
\thispagestyle{empty}
\begin{center}
{\LARGE \bf
On Geometric Observables in Gravity
}

\bigskip\vspace{1cm}
\begin{center}{\large Manthos Karydas${}^{a}$, Murat Kolo\u{g}lu${}^{b,c}$, Allic Sivaramakrishnan${}^{b,d}$}\end{center}
%\vspace{3mm}
{
\it ${}^a$Physics Division, Lawrence Berkeley National Laboratory, 
Berkeley, CA\\
${}^b$Department of Physics, Yale University, New Haven, CT\\
${}^c$School of Mathematical Sciences, Queen Mary University of London,\\
London, United Kingdom\\
${}^d$
Department of Physics and Astronomy, Purdue University,\\ West Lafayette, IN
}

 \end{center}
 
\bigskip

\centerline{\large\bf Abstract}
\begin{quote} \small

Measures of geometry like distance and time are staples of classical gravity. While geometric quantities arise broadly in quantum gravity, their graviton corrections are less explored. In this paper, we study geometric observables in perturbative quantum gravity. We form distance and time observables that model measurements, including in the presence of spacetime fluctuations. Gravitational dressing ensures these constructions are coordinate invariant. Then we show how these finite-distance observables become memory and Shapiro time delay operators upon taking asymptotic limits. We compute leading graviton effects in correlators of geometric observables throughout. Altogether, our results identify a class of geometric observables that link experiments and operations on the celestial sphere.

\end{quote}
	
\setcounter{footnote}{0}

\setcounter{tocdepth}{3}
\newpage
\tableofcontents

\newpage

\section{Introduction}

Spacetime fluctuations at different scales usually warrant very different observables. Boundary quantities are prime candidates for observables in the exotic, high-energy regime of quantum gravity. When spacetime froth obliterates locality at high energies, retreating to a non-gravitational boundary can allow one to safely make measurements. At low energies, however, experiments carried out in a lab appear well-localized despite being subject to dynamical gravity. It is also a highly testable fact that the lab does not reside at any asymptotic boundary. Laboratory observables can sometimes be approximated by their distant cousins, at least when a separation of scales is present. This is famously the case for gravitational wave detectors like LIGO and detectors at particle colliders like the LHC. But in quantum gravity, the lab itself is adrift atop spacetime fluctuations. This unavoidable loss of locality, and with it scale, makes precise connections between bulk and boundary observables harder to discern. To make headway, we may need to navigate like early sailors, who combated their own watery delocalization by measuring locally and comparing to landmarks in the sky.

Measures of distance and time are fundamental in classical gravity, which makes them appealing targets for observables in quantum gravity both locally and at asymptotic infinity. These geometric quantities are also relatively simple, primal versions of widely studied geometric observables like area, focusing, and geodesic deviation. Distance and time observables are central to the study of gravitational memory and causality, and belong to a web of other formal and experimental topics as well. As we will aim to understand these topics better in the low-energy effective field theory (EFT) of quantum gravity, we briefly give some relevant context.

Gravitational memory is a classical phenomenon, for example the permanent residual displacement of masses left behind by a passing gravitational wave. Laser interferometers are high-precision tools for directly measuring distances \cite{aasi2015advanced, maggiore2008gravitational}, but measuring memory with interferometers is more challenging in existing experiments than the detection of gravitational waves, for example see  \cite{Mitman:2024uss}. In the simplest models, the interferometer observable is a photon's round trip time, even though it is sometimes quoted as a distance, since the two are equivalent for low-frequency gravitational waves  \cite{aasi2015advanced,ligo-note}. See \cite{Domcke:2024abs} for a corresponding study of high-frequency gravitational waves. Quantum gravity signals may entail high-frequency effects, however, which could require distinguishing distance and time observables. Various proposals for quantum gravity signals in interferometers do indeed invoke distance and round-trip time observables \cite{Parikh:2020fhy, Parikh:2020kfh, Parikh:2020nrd, Verlinde:2019xfb, Kanno:2020usf, Fransen:2025npa, Aalsma:2025bcg, Ciambelli:2026pwi}. How these compare to the fully relativistic distance and time observables built using EFT methods remains somewhat opaque, and exploration would benefit from a more detailed understanding of perturbative corrections to the latter. Having technical control over such observables in effective field theory may also clarify connections between various formal and experimental topics, including celestial holography \cite{Raclariu:2021zjz, Pasterski:2021raf}, quantum gravity signatures in the GQuEST experiment \cite{Verlinde:2019xfb, Verlinde:2019ade, Aalsma:2025bcg, Zurek:2020ukz, Vermeulen:2024vgl}, memory correlators \cite{Moult:2025njc, Oertel:2026wsm, Oertel:2026oqv}, and more broadly the study of infrared behavior of gauge theories and gravity \cite{weinberg1964photons,weinberg1965photons,Strominger:2017zoo, WeinbergPhysRev.140.B516}. 

Causality is a well-tested pillar of modern physics and a workhorse for a number of landmark theorems and constraints in gravity. One useful form of causality is that the arrival time of a null ray at an asymptotic boundary must obey certain conditions \cite{Gao:2000ga}. Positivity of this Shapiro time delay has been used to place bounds on EFT coefficients in gravity, including the graviton three-point vertex in flat space and in AdS \cite{Camanho:2014apa, Hartman:2015lfa, Afkhami-Jeddi:2017rmx}. Time delays can be defined geometrically or extracted from the phase of a scattering process with eikonal kinematics. But for general kinematics and more broadly in quantum gravity, it is less clear how to define such a time delay, let alone whether it might be positive or even bounded in some rigorous sense. This has implications for the fate of light cones in quantum gravity. A number of widely leveraged relationships link causality and other deep properties of gravity, quantum field theory, and quantum information. In particular, time delays connect to focusing and the averaged null energy condition \cite{Hartman:2022njz}, which in turn connects to more general properties of light ray operators and energy correlators \cite{Kravchuk:2018htv, Caron-Huot:2022eqs, Herrmann:2024yai, Moult:2025nhu}. These one-way time delays are however less studied as bulk observables compared to the round-trip time delays of interferometers, even though the latter type is more involved since it includes a reflection.

Geometric quantities appear widely in quantum gravity. Despite this, explicit computations of their graviton corrections remain challenging, and the subject of active research. Even the landscape of graviton corrections to distance and time observables in linearized quantum gravity remains largely uncharted, although see \cite{ Ford:1994cr, Ford:1997zb, Yu:2009mv, Khavkine:2011kj,Bonga:2013uha, Vieira:2018bjn, Donoghue:2021meq, Carney:2024wnp, Freidel:2026hed, Freidel:2026gcp}. Previous work on graviton corrections in a fully relativistic setting has mainly focused on individual observables, often differing in their choices of gauge, kinematic regimes, experimental application, and treatment of diffeomorphism invariance. In this paper, we advance this direction by showing that a number of these geometric observables across different distance scales can be derived from a single, common origin both in classical and linearized quantum gravity. This is supported by explicit computations of leading graviton corrections, including to vacuum correlators. To construct graviton corrections to geometric observables, we use modern constructions of dressed or relational observables \cite{Giddings:2025xym}. This approach makes diffeomorphism invariance of expressions manifest, easily verifiable, and furnishes ready-made algorithms for computing perturbative corrections in EFT \cite{Cheung:2026euf}. Working in flat space, we calculate graviton corrections to an array of geometric observables, explore their interrelations, and show how they interpolate between experiments in the bulk and observables on the celestial sphere. 

We emphasize that while some of our computations are new, other parts of our results are essentially known or anticipated in the gravity literature, even if not explicitly stated or packaged into the expressions we give. Our main, specific contribution is to connect bulk and boundary geometric observables from start to finish, and show that this includes the graviton corrections that might make this task subtle to begin with.

We now give an outline of the paper and our main results. In Section \ref{sec:geometric_observables}, we give manifestly diffeomorphism-invariant proper distance and time observables accessible to bulk observers, which we model by geodesics. We then compute leading-order quantum gravity corrections in gravitational EFT \cite{Donoghue:1995cz, Donoghue:2022eay, Burgess:2003jk}. We comment on how these observables compare and are related to more realistic experimental setups.

In Section \ref{sec:asymptotic_limit}, we then take asymptotic limits of the bulk geometric observables. Putting our distance observable in Bondi coordinates allows us to explicitly match onto the asymptotic shear tensor usually quoted in Bondi gauge. The arrival time delay observable becomes the Shapiro time delay. This gives a geometric definition of the Shapiro time delay as an operator in quantum gravity and an avenue to compute higher-order corrections. 

In Section  \ref{sec:asymptotic_observables}, we compute the asymptotic counterparts of bulk observables that we identified. Specifically, we calculate leading graviton contributions to correlators and matrix elements of Shapiro time delays and memory. We discuss the infrared behavior of the round-trip time delay observable. We encounter IR divergences but ultimately identify IR-finite observables.

In Appendix \ref{app:geometric_observables_arbitrary} we give a compact derivation of all the classical geometric observables we study from a common starting point, without choosing a gauge, and demonstrate their diffeomorphism invariance using EFT techniques. In Appendix \ref{app:geometric_observables_synchronous}, we work out the same observables in the more familiar synchronous gauge and then compute their vacuum correlators. While some of the results are directly relevant to our main results, and may be of independent interest, we omit these details from the body to streamline presentation.

\section{Geometric observables at finite distance}
\label{sec:geometric_observables}

There are several standard ways to describe how far apart two timelike geodesics are \cite{Misner:1973prb}. One is the expansion $\theta$ of timelike congruences. This is a scalar, although it is a local measure and is not itself a distance. Another is geodesic deviation $\xi^\mu$, which captures the acceleration of two nearby geodesics. This is a spacetime vector and it captures only distances between infinitesimally nearby geodesics, although see \cite{Kim:2025jyy} for recent work on extension to finite separation. Note that quantum gravity extensions of area and focusing have received significant attention for their role in, for example, quantum focusing \cite{Bousso:2015mna} and quantum extremality \cite{Colin-Ellerin:2025dgq}. 

A more direct measure of distance is the length of a spacelike geodesic. If the endpoints are specified in an invariant way, the geodesic length is diffeomorphism invariant. However, an experimentalist cannot traverse a spacelike curve to directly measure its length. By contrast, the times measured by sending and receiving null rays can more directly be measured by an experimentalist, and are common proxies for the distance between the two geodesics. In this section, we will uplift several of these geometric observables to linearized quantum gravity and unpack their differences. 

Our approach uses the ideas in \cite{Carney:2024wnp, Sivaramakrishnan:2025srr, Cheung:2026euf}, which are well-utilized in general relativity. We consider perturbations around flat space, $g_{\mu\nu} = \eta_{\mu\nu}+\delta g_{\mu\nu}$, with $\delta g_{\mu\nu} = \kappa h_{\mu\nu}$, and $\kappa = \tilde{\ell}_p = \sqrt{32\pi G_N}$, the reduced Planck length. In this normalization, we perform perturbation theory in $\delta g$ and $h$ is dimensionful. We work in mostly-plus signature, $\eta_{\mu\nu} = \text{diag}(-,+,+,\cdots)$, in units in which $\hbar=c=1$, and we perturb about $d$-dimensional Minkowski spacetime.

We will consider two worldlines, with coordinates $x_1, x_2$, that perturbatively obey the geodesic equation,
\begin{equation}
\ddot{x}_i^\mu + \Gamma^\mu_{\rho\sigma} \dot{x}_i^\rho \dot{x}_i^\sigma = 0, ~~~~~~~~ x_i = x_i^{(0)}+ x_i^{(1)}+x^{(2)}_i+\cdots\,,
\end{equation}
where the superscript denotes the perturbative order in $\delta g$. The worldlines are initialized in the far past and parametrized by their proper times. These worldlines will model observers. Concretely, we suppose 
\begin{align}
& x_2^{(0),\mu} - x_1^{(0),\mu} = L n^\mu \equiv L (0,\mathbf{n}),
\\
& \dot{x}_1^{(0),\mu}  = \dot{x}_2^{(0),\mu} = \delta^{\mu}_{0},
\end{align}
so that the unperturbed geodesics  are stationary and separated by $L$ in a direction $n^\mu$ with $n = (0, \mathbf{n})$, where $\mathbf{n}$ is a unit spatial vector, and we use bold notation for spatial vectors. See for example \cite{Donnelly:2015hta, Frob:2023awn, Cheung:2026euf}, and Appendix \ref{app:geometric_observables_arbitrary_diffs_and_initializations} for further details of the setup.

We can form geometric observables $O_g$ that measure distances and times associated with geodesics that connect the two observer worldlines. We will find that $O_g =  O^{(0)}_g + O^{(1)}_g + \cdots$ can be written in linearized gravity as
\begin{equation}
O_g^{(1)}(\tau) = -L \int_{-\infty}^\tau d\tau' (\tau-\tau') \int_{\gamma(\alpha)} d\alpha \, R^{(1)}_{0n0n}(x^{(0)}(\tau',\alpha)),
\label{OgArbitraryGauge}
\end{equation}
where above $\alpha$ is dimensionless and parametrizes the relevant unperturbed geodesic $\gamma(\alpha)$. This geodesic is coordinatized by $x^\mu(\tau',\alpha)$, has a perturbative expansion in $\delta g$ as $x = x^{(0)}+x^{(1)}+\ldots $, and can be a spacelike, null, or timelike geodesic. The expression in Eq. \eqref{OgArbitraryGauge} is valid in any gauge. Diffeomorphism invariance of all such $O_g$ is immediate, as the linearized Riemann tensor is diffeomorphism-invariant, though for a more detailed analysis see Appendix  \ref{app:geometric_observables_arbitrary_diffs_and_initializations}.  The form Eq. \eqref{OgArbitraryGauge} clarifies how geometric observables are related to the more familiar dressed local operators: they are a gravitationally dressed Riemann tensor integrated over the geodesic curve $\gamma$.

The synchronous gauge expression we use for computations is the simpler quantity
\begin{align}
 O_g^{(1)}(\tau) =&  \kappa\frac{L}{2}\int_{\gamma(\alpha)} d\alpha \, h_{nn}(x^{(0)}(\tau,\alpha)). && (\text{Synchronous gauge})
\label{OgSynchronousGauge}
\end{align}
While the gauge-independent form involves several integrals, in Eq. \eqref{OgSynchronousGauge} we see that $O_g^{(1)}$ in synchronous gauge are simple geodesic integrals of $h$. 

Geometric observables furnish rulers, clocks, and so on that describe spacetime locations in a diffeomorphism-invariant way.\footnote{These are not to be confused with clock and ruler fields, which are field theory degrees of freedom that model physical clocks and rulers. Just as classical geometry is the purely geometric data underlying realistic measurements by corporeal devices, here we study the similarly pure quantum geometric data.} This diffeomorphism invariance comes at the expense of locality, specifically the $\alpha$ integral of $x(\tau,\alpha)$. Here, the Christoffel symbol $\Gamma_{\rho\sigma}^\mu = 
\Gamma^{(1),\mu}_{\rho\sigma}+ \ldots$ and Riemann tensor $R_{\mu\nu\rho\sigma} = R^{(1)}_{\mu\nu\rho\sigma}+\ldots$ are given by their linearized versions, and we use similar notation for other curvature quantities.\footnote{Christoffel symbols in flat space, $\Gamma^{(0),\mu}_{\rho\sigma}$, can be nonzero for some coordinate systems, for example in curvilinear sets of coordinates. For simplicity, here we imagine an affine choice of coordinates with $\Gamma^{(0),\mu}_{\rho\sigma}=0$. Of course, the Riemann tensor is always zero in flat space regardless of the choice of coordinates.} We denote contractions with the unperturbed worldline velocities $\dot{x}_i = (1,0,0,0)$ using the compact notation $\dot{x}^{(0)}_{i,\mu} v^\mu = v_0$, and similarly for contractions with $n$, so that compact expressions like the following are Lorentz scalars: $R^{(1)}_{0n0n} = R^{(1)}_{\mu\nu\rho\sigma} \dot{x}_i^\mu n^\nu \dot{x}_i^\rho n^\sigma$. 

A number of geometric quantities take a similar form in linearized gravity:
\begin{align}
&\text{Geodesic motion}  &  [x^\mu] &= \text{Length} &   \ddot{x}_i^{(1),\mu} + \Gamma^{(1),\mu}_{00} &= 0,
\\
&\text{Geodesic focusing} &  [\theta] &= \text{Length}^{-1} & \ddot{\theta}^{(1)} + \dot{R}_{00}^{(1)} &= 0,
\\
&\text{Geodesic deviation} & [\xi^\mu] &= \text{Length} &   \ddot{\xi}^{(1),\mu}+\partial_\nu \Gamma ^{(1),\mu}_{0 0} \xi^{(0),\nu} &= 0,
\\
&\text{Geometric observables} & [O_g] &= \text{Length} & \ddot{O}_g^{(1)}(\tau)+ L\int_{\gamma(\alpha)} d\alpha  \, R^{(1)}_{0n0n}(x^{(0)}(\tau,\alpha)) &= 0.
\end{align}
To compare various geometric measures, we have displayed the second-order differential equations that govern the evolution of focusing and geodesic deviation at $\mathcal{O}(\kappa)$. Here we expanded the expansion scalar $\theta = \theta^{(0)}+\theta^{(1)}+\cdots$  and geodesic deviation vector $\xi^\mu = \xi^{(0),\mu}+\xi^{(1),\mu}+\cdots$. $R_{\mu\nu}$ is the Ricci tensor. In the list above, the geometric observables $O_g^{(1)}$ are the only scalars with dimensions of length. The general formula we find for observables $O^{(1)}_g$ shows they can obey simple second-order differential equations more typical of dynamical objects with independent definitions. That is, these can be thought of as equations of motion for idealized clocks and rulers. Whether such structure is more general remains unclear, but this simplicity makes our computations particularly straightforward in quantum gravity.

We will take a standard approach to promoting classical quantities to observables in quantum gravity in anticipation of performing perturbative calculations: we will compute for off-shell $h$ and then perform the path integral over $h$. This is equivalent to canonically quantizing $h$, though the distinction will not be important here. We will derive expressions for $O_g^{(1)}$ valid for $h$ off shell, as required. We will analyse and compute vacuum correlators in synchronous gauge,
\begin{equation}
\braket{0|O_g^{(1)}(\tau_1) O_g^{(1)}(\tau_2)|0} = \frac{\kappa^2L^2}{4}\int d\alpha_1 \int d\alpha_2 \braket{0|h_{nn}(x^{(0)}(\tau_1,\alpha_1))h_{nn}(x^{(0)}(\tau_2,\alpha_2))|0},
\end{equation}
as well as matrix elements of the form,
\begin{equation}
\braket{0|O_g^{(1)}(\tau)|h(k)} = \frac{\kappa L}{2} P_{n\rho}P_{n\sigma} \epsilon^{\rho \sigma} \int d\alpha \, e^{ik x^{(0)}(\tau,\alpha)},
\label{OgGravitonMatrixElementSynchronousGauge}
\end{equation}
with the projector $P_{\mu\rho} = \eta_{\mu\rho} - \frac{k_\mu \eta_{0\rho}}{k_0-i\varepsilon}$. See Appendix \ref{app:geometric_observables_arbitrary_diffs_and_initializations} for further details. 

The expression in Eq. \eqref{OgGravitonMatrixElementSynchronousGauge} may appear opaque, but by expanding out the projectors, it becomes apparent we can attribute diagrammatic interpretations to the different terms \cite{Cheung:2026euf}. The $\epsilon_{nn}$ term comes with no powers of $1/(k^0-i\varepsilon)$. Inspection of Eq. \eqref{OgArbitraryGauge} shows this occurs because the $\partial_\tau^2$ term cancelled the worldline propagator. This term can be interpreted as the graviton changing $O_g$ by landing on the geodesic $x(\tau,\alpha)$. This is precisely the naive contraction of $O_g$ in synchronous gauge, as in Eq. \eqref{OgSynchronousGauge}, with an external graviton. However, diffeomorphism invariance requires that this is not the only contribution. The other terms in Eq. \eqref{OgGravitonMatrixElementSynchronousGauge} involve factors of $1/(k^0-i\varepsilon)$ and can be viewed as arising from the worldline dressing. The presence of the denominator $1/(k^0-i\varepsilon)$ in these terms captures how the graviton alters $O_g$ by landing on the observer worldlines $x_i$. Such interpretations are expected to persist in more complex quantities as an unavoidable consequence of diffeomorphism invariance \cite{Cheung:2026euf}.

\subsection{Proper distance}
\label{subsec:proper_distance}

We begin by defining a manifestly diffeomorphism-invariant form of proper distance between two worldlines. Then we compute the leading correction to its two-point function in linearized quantum gravity.

Consider two points, $x_1(\tau_1), x_2(\tau_2)$ specified by the proper times $\tau_1, \tau_2$ elapsed from the initialization surface. For simplicity, we choose $\tau_1=\tau_2 \equiv \tau$ such that the points $x_1^{(0)}(\tau), x_2^{(0)}(\tau)$ are spacelike separated. Then, as we show in Appendix \ref{app:proper_distance_arbitrary_gauge}, proper distance $\sigma(\tau)$ is 
\begin{equation}
\sigma(\tau) =L-L\int_{-\infty}^{\tau} d\tau'(\tau-\tau') \int d\lambda \, R^{(1)}_{0n0n}(x^{(0)}(\tau',\lambda)),
\end{equation}
where $x^{(0)}$ is the unperturbed geodesic between $x_1(\tau)$ and $x_2(\tau)$, parametrized by $\lambda$. 

\begin{figure}[h]%changed from [t] to [h] so it'd appear where it's written.
\centering
\begin{tikzpicture}[scale=0.7]

% Thick vertical worldlines for L and R
\draw [very thick] (3,0) -- (3,7);
\draw [very thick] (6,0) -- (6,7);

% Labels L and R below
\node at (3,-0.5) {$1$};
\node at (6,-0.5) {$2$};

% Arrows pointing up at the base of each worldline
\draw [->, thick] (3,0) -- (3,1);
\draw [->, thick] (6,0) -- (6,1);

% Distance label L between the two worldlines
\draw [<->] (3.3,1.5) -- (5.7,1.5);
\node at (4.5,1.8) {L};

% Points on the worldlines
\fill (3,4.5) circle (3pt);
\fill (6,4) circle (3pt);

% Labels for the points
\node [left] at (2.8,4.5) {$x_1(t_1)$};
\node [right] at (6.2,4) {$x_2(t_2)$};

% wiggly geodesic connecting the two points
\draw [thick, domain=0:1, smooth, variable=\x, samples=200] 
    plot ({3 + 3*\x}, {4.5 - 0.5*\x + 0.12*sin(8*\x*180)});

% Sigma label above the geodesic
\node at (4.5,5.0) {$\sigma(t_1, t_2)$};

% Coordinate axes (bottom-left)
\draw [->] (0,1.5) -- (1.5,1.5);
\draw [->] (0,1.5) -- (0,3);
\draw [->] (0,1.5) -- (-0.5,0.5);

% Axis labels
\node [right] at (1.5,1.5) {$z$};
\node [left] at (0,3) {$t$};
\node [left] at (-0.5,0.5) {$x^i$};

\end{tikzpicture}
\caption{The $x_1, x_2$ geodesics. The wavy curve is a spacelike geodesic between the points $x_{1}(\tau_1)=(t_1,\mb{x}_1)$ and $x_{2}(\tau_2)=(t_2,\mb{x}_2)$ on the two timelike geodesics with  proper distance $\sigma(t_{1},t_{2})$. Proper times $\tau_1, \tau_2$ specify the endpoints. In the above, we have used synchronous gauge for the metric, $ds^{2}=- dt^2 + g_{ij}dx^{i}dx^{j}$, for which the timelike geodesics remain straight lines. In computations, we work in four spacetime dimensions and take the worldline separation to lie in the $z$ direction for definiteness.}
    \label{fig:geodesics}
\end{figure}
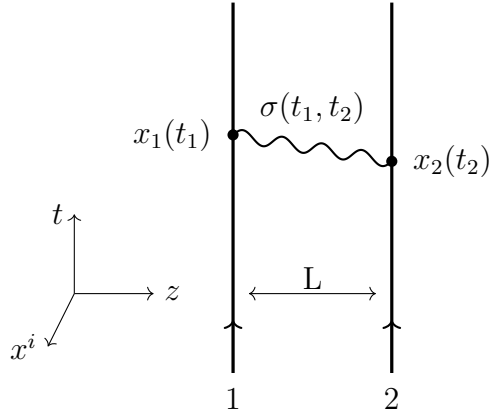
This obeys the second-order equation,
\begin{equation}
\frac{d^2}{d\tau^2}\sigma(\tau) =-L\int d\lambda \, R^{(1)}_{0n0n}(x^{(0)}(\tau,\lambda)).
\end{equation}
The ``acceleration'' of the distance $\sigma$ between the two geodesics is just the curvature integrated over the spacelike geodesic between them. 

In synchronous gauge, $h_{0\mu} = 0$ and $R^{(1)}_{0i0j} = -\frac{\kappa }{2} \partial_t^2 h_{ij}$, so $\sigma$ simplifies to 
\begin{equation}
\sigma(\tau) = L+\kappa\frac{L}{2}\int d\lambda \, h_{nn}(x^{(0)}(\tau,\lambda)).
\end{equation}
We derive this more familiar expression directly in synchronous gauge in Appendix \ref{app:proper_distance_synchronous}. 
To perform computations, we will work in $d=4$ spacetime dimensions and take $n$ to point in the $z$ direction without loss of generality. The vacuum Wightman function is 
\begin{equation}
\label{two_point_distance_corellator_hzz}
\braket{\sigma(t)\sigma(0)}_c = \frac{\kappa^2 L^2}{4} \int_{0}^{1} d\lambda d\lambda' \braket{h_{zz}(
x^{(0)}(t,\lambda))h_{zz}(
x^{(0)}(0,\lambda'))},
\end{equation}
which is obtained via analytic continuation by taking the $\varepsilon\rightarrow 0$ limit of $t\rightarrow t - i\varepsilon$. In the above Wightman function, the $h_{zz}$ are on shell, and so 
%in synchronous gauge can also satisfy transverse-traceless (TT) gauge as well.
we can use the residual gauge freedom of synchronous gauge to also impose transverse-traceless (TT) gauge conditions on $h_{ij}$.\footnote{For extensions to higher-point correlators, see \cite{Cheung:2022pdk, Meltzer:2020qbr} for discussion of on-shell correlation functions.} We can therefore use the TT gauge two-point function to obtain
\begin{equation}
\label{sigma_sigma_correlator_PSD_form}
\braket{\sigma(t-i\varepsilon)\sigma(0)}_c \equiv \int_{0}^{\infty} dk e^{-i k t - k \varepsilon} S_{\sigma \sigma}(k),
\end{equation}
with noise
power spectral density (PSD) $S_{\sigma\sigma}$ given by
\begin{equation}
S_{\sigma\sigma}(k)= \frac{\kappa^2}{8\pi^2 k} \bigg( -8/3+\left(1-2/(k^2L^2)\right)\cos(kL)
+\frac{2 \sin(kL)}{k^3 L^3} 
+ \frac{\sin(k L )}{k L} + k L\, \text{Si}(kL)
\bigg),
\label{sigmasigmapsd}
\end{equation}
where $\text{Si}(x) = \int_0^x dt \sin(t)/t $ and $\kappa \sim 10^{-34}$ m, the reduced Planck length. As  $S_{\sigma\sigma }(k) \sim \kappa^2 L^2 k$ near $k = 0$, the correlator $\braket{\sigma(t-i\varepsilon)\sigma(0)}_c$ has no IR divergence. Since $S_{\sigma\sigma}(k) \sim \kappa^2L  $ at large $k$, there naively appears to be a UV divergence, but as explained in \cite{Carney:2024wnp}, we must hold $\varepsilon \neq 0$ while performing the integral and only afterwards send $\varepsilon \rightarrow 0^{+}$, and this becomes finite when integrating over a finite-time smearing window with a filter function, as this contributes an additional $1/k^2$ suppression at large $k$.

Using the $i\varepsilon$ definition of this correlator, we can remain at finite $i\varepsilon$ to compute the commutator and anticommutator from $\braket{[\phi(x_1), \phi(x_2)]}=2i\,\text{Im}\braket{\phi(x_1) \phi(x_2)}$ and $\braket{\{\phi(x_1), \phi(x_2)\}}=2\,\text{Re}\braket{\phi(x_1) \phi(x_2)}$.

For $t>L$ we find that the commutator vanishes, as expected from the causality properties of the underlying graviton two-point function. We find the anticommutator is   
\begin{align}
\braket{ \{ \sigma(t), \sigma(0)\}}_c =
&
\frac{\kappa^{2}}{72\pi^2 s}\bigg( 
34 s -6 s^3 -3(s-1)^3(3+s)\log(s-1)
\nonumber
\\
&
+48 s \log(s) + 3(s-3)(1+s)^3 \log(1+s)
\bigg),\quad s>1,
\end{align}
 where we defined the dimensionless quantity $s = t/L$. For $0<t<L$, both the commutator and anticommutator are non zero,
 \begin{align}
\braket{ \{ \sigma(t), \sigma(0)\}}_c =
&
\frac{\kappa^{2}}{72\pi^2 s}\bigg( 
34 s -6 s^3 -3(s-1)^3(3+s)\log(1-s)
\nonumber
\\
&
+48 s \log(s) + 3(s-3)(1+s)^3 \log(1+s)
\bigg),\quad 0<s<1,
\end{align}
and
\begin{equation}
\braket{[\sigma(t),\sigma(0)]}
= i\frac{\kappa^{2}}{24\pi s} (s-1)^3 (3 +s),\quad 0<s<1\,.
\end{equation}
As this is nonzero, it confirms that this definition of distance is a quantum operator. While the expression above approaches a non-zero value at $t=0$, one can restore the contact terms relevant to the $t=0$ case, and doing so confirms the commutator does indeed vanish at $t=0$ as expected. See  Appendix \ref{app:proper_distance_synchronous} for further details.

Finally, we can compute the one-graviton matrix element,
\begin{align}
\braket{0|\sigma(t)|h(k)} &= \frac{\kappa L }{2} P_{z\rho}P_{z\sigma} \epsilon^{\rho \sigma} \int d\lambda e^{ik^\mu x^{(0)}_\mu(\tau,\lambda)}
\nonumber
\\
&
=
\frac{\kappa L }{2} P_{z\rho}P_{z\sigma} \epsilon^{\rho \sigma} \int d\lambda e^{i(-k^0 t+\lambda k^z L )}
\nonumber
\\
&=
\frac{\kappa L }{2} P_{z\rho}P_{z\sigma} \epsilon^{\rho \sigma} e^{-i k^0 t+i k^z L/2}\frac{\sin(k^z L /2)}{k^z L/2}
,
\end{align}
where we used the straight line trajectory $x^{(0),\mu}(\tau,\lambda) = (t,0,0,\lambda L)$, as described in Appendix \ref{app:proper_distance_synchronous}. This is finite, but as explained in \cite{Cheung:2026euf}, can in principle acquire singularities for soft $k$ when the two worldlines are taken to have different velocities.

\subsection{Arrival time}
\label{sec:arrival_time}
The arrival time of a null ray is a fundamental timing observable. It underlies certain clock synchronization protocols and provides a notion of timing that arises from the causality of the underlying spacetime. In this section, we extend this quantity to linearized quantum gravity. 

We consider a single photon that is emitted from worldline $x_1$ and arrives at worldline $x_2$ some time later. This setup describes a bulk or finite-distance version of the classic Shapiro time delay in a fully gauge-invariant way. There are several related time delay observables that employ this setup, which we analyse in further detail in Appendix \ref{app:arrival_time_arbitrary_gauge}. The upshot is the following. Consider a photon emitted at $x_1(\tau_1)$ and arriving at $x_2(\tau_2)$. It is tempting to identify $\tau_2$ as an arrival-time observable. However, as only relative times are measurable, the value of $\tau_2$ will depend on the timing of some prior event. For simplicity, we have been initializing the worldline at past infinity. But then $\tau_1, \tau_2 = \infty$ in this setup, which means that the photon arrival observable is not well-defined.

To form a good observable from $\tau_2$, we must subtract off the negative infinite initial time in a diffeomorphism-invariant way. The quantity $\tau_2-\tau_1$ is an observable of this type. If the observers cross paths and synchronize clocks, the observer $x_2$ can measure $\tau_2-\tau_1$. We will begin with this simple observable and then move to one measurable along $x_2$ without the need for prior synchronization. As we show in Appendix \ref{app:arrival_time_arbitrary_gauge}, $\tau_2-\tau_1$ is
\begin{equation}
\tau_2-\tau_1 = L-
\int_0^L d\eta   
\int^{t_1^{(0)}+\eta}_{-\infty} dt' (t_1^{(0)}+\eta-t') R^{(1)}_{0n0n},
\end{equation}
which in synchronous gauge reduces to the familiar form of the one-direction time delay,
\begin{equation}
 \tau_2-\tau_1=L+\frac{\kappa}{2}
\int_0^L d\eta \,   
h_{nn}(x(\eta)).
\end{equation}
See \ref{app:arrival_time_synchronous} for an independent derivation of this more familiar synchronous gauge expression.

We will proceed in light cone coordinates to clarify the connection with other well-known objects and calculations in the literature. Consider a light ray propagating in the $z$ direction, with the worldlines separated along $n = (0,0,0,1)$. We use the light cone coordinate 
\begin{equation}
u\equiv t- z.
\end{equation}
Using the resulting metric, Eq. \eqref{metric_synchronous_gauge}, we perturbatively expand the light-cone coordinate of the light ray as
\begin{equation}
\label{light_cone_coordinate}
u(\eta)= u^{(0)}(\eta)+ u^{(1)}(\eta)+ \dots ,
\end{equation}
where $\eta$ is an affine parameter for the light ray. In what follows, we will consider a symmetric configuration of the worldlines, with $\eta \in [-T,T]$, where $T=L/2$, and place the unperturbed worldlines at $z= \pm T$. We will describe null rays initialized at null time $u_0$. We can then label the unperturbed null rays according to their values of $u$ as $x^{(0)}(\eta;u_0)$ with 
\begin{equation}
x^{(0)}(\eta ;u)= (\eta+u,0,0,\eta).
\label{unperturbed_rays}
\end{equation}
A straightforward computation (see Appendix \ref{app:arrival_time_synchronous}) then gives the null time at $\eta$ as
\begin{equation}
u(\eta;u_0)= u_{0}+\frac{\kappa}{2}\int_{-T}^{\eta}d\eta' h_{zz} (x^{(0)}(\eta';u_{0}))+ O(h^2).
\label{vshift}
\end{equation} 
In light cone coordinates, 
\begin{equation}
\tau_2-\tau_1 = L + u(T;u_0)-u(-T;u_0) \equiv L+u(u_0),
\end{equation}
where for compactness we define $u(u_0) \equiv u(T;u_0)-u(-T;u_0)$. The presence of metric perturbations causes fluctuations on the light-cone time $u$ of the light ray, as shown in Fig. \ref{fig:light-rays}. The coordinate values of $u$ can contain an infinite contribution if we take an initialization surface at past infinity to be their zero value, and this infinity has been subtracted off in $u$, or equivalently in $\tau_2-\tau_1$. Taking two null rays emitted a time $u_0'-u_0 = t$ apart, the leading-order graviton contribution to the connected two-point Wightman function of \eqref{vshift} is, expressed in synchronous gauge,
\begin{align}
\braket{(\tau_2(u_0)-\tau_1(u_0))(\tau_2(u_0')-\tau_1(u_0'))}_c
=&
\braket{ u(u_{0}) u(u_{0}')}_c
\nonumber
\\
=&
\frac{\kappa^2}{4}\int_{-T}^{T} d\eta d\eta' \braket{h_{zz}(x^{(0)}(\eta; u_{0}))h_{zz}(x^{(0)}(\eta'; u'_{0}))},
\label{uu_wightman_function}
\end{align}
which can be evaluated. See \cite{Sivaramakrishnan:2025srr} for explicit expressions.

We can also compute the matrix element with an external graviton.
\begin{align}
\braket{0|\tau_2-\tau_1|h(k)} &= \frac{\kappa }{2} P_{z\rho}P_{z\sigma} \epsilon^{\rho \sigma} \int d\eta e^{ik^\mu x^{(0)}_\mu(\tau,\eta)}
\nonumber
\\
&
=
\frac{\kappa }{2} P_{z\rho}P_{z\sigma} \epsilon^{\rho \sigma} \int_0^L d\eta e^{i(-k^0 (\eta+t) + k^z \eta   )}
\nonumber
\\
&=
\frac{\kappa L }{2} P_{z\rho}P_{z\sigma} \epsilon^{\rho \sigma} e^{-i k^0 t+i(k^z-k^0)L/2}\frac{\sin((k^0-k^z)L/2)}{(k^0-k^z)L/2}
,
\end{align}
where in the above we used $x^{(0),\mu}(\tau,\eta) = (t+\eta,0,0,\eta)$ with $\eta \in [0,L]$. This matrix element is finite. Here $k$ can become collinear with the null ray, but this does not generate a singularity.

\subsection{Round trip time}

The observable in a laser interferometer is often quoted either as the proper distance $\sigma$ between beamsplitter and mirror, or round trip time $\tau$ of a bouncing photon  \cite{Misner:1973prb,maggiore2008gravitational, Rakhmanov:2004eh}. For gravitational wave detection, when the incident gravitational wave has wavelength $\lambda_{GW} \gg L$, with interferometer arm length $L$, these two quantities are simply related, $\tau=2\sigma$, at leading order in the $L/\lambda_{GW}$ expansion. However, outside of this large-wavelength regime, the observables $\tau$ and $\sigma$ are in general different. Here, we revisit, compare, and discuss the differences, including in linearized quantum gravity.

The round trip proper time turns out to take an exceedingly simple form to linear order in $h$, independent of the gauge chosen. For a null ray that is emitted at $x_1(\tau_1)$, reflected at $x_2(\tau_2)$ and received at $x_1(\tau_3)$, the elapsed proper time $\tau \equiv \tau_3-\tau_1$ can be written to leading order in $h$ in terms of the independently gauge-invariant one-way time observables we already identified,
\begin{equation}
\tau = (\tau_3-\tau_2)+(\tau_2-\tau_1),
\end{equation}
where
\begin{align}
\tau_2-\tau_1 &=L
-
\int_0^L d\eta   
\int^{t_1^{(0)}+\eta}_{-\infty} dt' (t_1^{(0)}+\eta-t') R^{(1)}_{0n0n},
\nonumber
\\
\tau_3-\tau_2 &=L
-
\int_0^L d\eta   
\int^{t_2^{(0)}+\eta}_{-\infty} dt' (t_2^{(0)}+\eta-t') R^{(1)}_{0n0n}.
\end{align}
The light ray here is initialized at time $\tau_1$, which is captured by coordinate time $t_1$. The time the light ray bounces is $t_2$, and is related at zeroth order as $t_2^{(0)} = t_1^{(0)}+L$. The paths integrated over are the outgoing and then ingoing unperturbed null rays for $\tau_2-\tau_1$ and $ \tau_3-\tau_2$ respectively. We derive this result in Appendix \ref{app:arrival_time_arbitrary_gauge}, to which we refer the reader for additional details and discussion.

In synchronous gauge, and with $n = (0,0,0,1)$ for definiteness, this simplifies to the familiar expression
\begin{equation}
 \tau =2L+ \frac{\kappa}{2}\int_{0}^{2L} d\eta\, h_{zz}(x^{(0)}(t,\eta)),
\end{equation}
where $x^{(0)}$ is the null ray leaving $x_1$, bouncing off $x_2$, and returning to $x_1$. For comparison, $\sigma$ is an integral of the same quantity, $h_{zz}$, but instead over a spacelike geodesic $x^{(0)}(t,\lambda)$. For the arbitrary gauge expressions, see Appendices \ref{app:proper_distance_arbitrary_gauge} and \ref{app:round-trip_time_arbitrary_gauge}. If $h_{zz}$ was a plane wave with $\lambda_{GW} \gg L$, then $h_{zz}$ is constant during this process and so $\tau=2\sigma$ at  $\mathcal{O}(h)$. However, unlike elapsed times, spacelike distances generically cannot be measured by a causal observer traversing the spacelike geodesic, and so can only be extracted from measurements that effectively reconstruct the full geometry. In quantum gravity the difference is more pronounced, as there is no classical geometry to reconstruct. One would have to instead perform measurements to reconstruct the quantum geometry, in other words, the quantum field $h_{\mu\nu}$.

The differences between proper distance and time are relevant for measurements of quantum gravity signatures in interferometers. In \cite{Carney:2024wnp}, the observable in a detailed model of a single-arm laser interferometer experiment (E) was computed,
\begin{align}
S_E(k) = &\kappa^2\frac{|\chi_c(k)|^2}{k}  \bigg( \cos(k L) - \frac{8}{3}+\frac{\sin(k L )}{k L} 
-\frac{2 \cos(k L)}{(k L)^2} + \frac{2 \sin(k L)}{(k L)^3}+(k L) \text{Si}(k L) \bigg),
\end{align}
up to an overall numerical coefficient, where $\chi_c$ is the response function that captures the geometry of the cavity. 

One might ask whether $S_E$ can be thought of in terms of a proper time or distance, like its classical counterparts. The proper time correlator computed in \cite{Carney:2024wnp} has the power spectral density (PSD)\footnote{We used the PSD definition of Eq.~\eqref{sigma_sigma_correlator_PSD_form}, which differs by a factor of $1/(2\pi)$ from the convention used in Ref.~\cite{Carney:2024wnp}. }
\begin{equation}
S_{\tau \tau}(k) = \frac{\kappa^{2}}{2\pi^2 k}
\bigg( \frac{3 - \cos(2 k L)}{6}  - \frac{3 + \cos(2 k L)}{2 k^2 L^2}
+ \frac{\sin(2 k L)}{k^3 L^3} \bigg).
\label{propertimetautauPSD}
\end{equation}
Clearly, $S_{E} \neq S_{\tau \tau}$. However, up to an overall numerical prefactor, $S_{E} = |\chi_c|^2 S_{\sigma \sigma}$, precisely the proper distance PSD  we computed in \eqref{sigmasigmapsd}. In other words, the detector observable computed in \cite{Carney:2024wnp} is evidently the proper distance we defined here, rather than proper time. Similarly, the work \cite{Parikh:2020fhy, Parikh:2020kfh, Parikh:2020nrd, Zurek:2020ukz} uses the separations of test masses as the experimental observable rather than proper time. It is worth understanding when $\tau$ as opposed to $\sigma$ is the relevant experimental observable, and why $\sigma$ appears here. Note that in \cite{Carney:2024wnp}, the optical readout mode is a standing wave in a cavity and is coupled to a spatial average of $h_{zz}$, which is essentially why the PSD is that of a proper distance. 

As expected, in the $k L \ll 1$ limit, the proper distance and time correlators agree: $S_{\tau\tau} = 4 S_{\sigma\sigma} = (2\kappa^{2} L^2 k)/(15 \pi^2)$.\footnote{The prefactor of $4$ in $S_{\tau\tau}=4 S_{\sigma\sigma}$ accounts for the proper time being defined as the round-trip time.} This is expected because $\braket{\tau(t) \tau(0)}$ and $\braket{\sigma(t,t)\sigma(0,0)}$ both behave for $t \gg L$ as $1/t^2$, which is also the large-$t$ behavior of $\braket{h_{zz}(0)h_{zz}(t)}$.

We leave exploration of how distance and time correlators enter into interferometer observables, including for GQuEST, to future work. We expect this can be understood by studying these quantities in the presence of high-frequency classical gravitational waves, including possibly above the optical frequency. For example, see \cite{Rakhmanov:2008is, Sharmila:2025lic, Laeuger:2026cve}.

\section{Taking the asymptotic limit}
\label{sec:asymptotic_limit}

So far, we have formed several observables that correspond to models of bulk experiments. We will now take suitable asymptotic limits. Matching local observables to asymptotic ones is notoriously subtle, including for widely studied quantities like collider observables in quantum field theory \cite{Moult:2025nhu}, the S-matrix \cite{Prabhu:2022zcr}, and related out-of-time-order cousins \cite{Caron-Huot:2023vxl, Caron-Huot:2025ymc}. 

In this section, we carefully study the asymptotic limit of the observables in Section \ref{sec:geometric_observables}. We have defined manifestly diffeomorphism-invariant quantities, but we will make contact with asymptotic limits commonly discussed in specific coordinate systems. We will need to account for the delta function support on the celestial sphere by a consistent $i\varepsilon$ prescription compatible with the asymptotic limit. Our goal here is to analyse the asymptotic limits of the metric perturbation $h_{\mu\nu}$, because all relevant boundary and bulk observables are functionals of $h_{\mu\nu}$. Quantizing $h_{\mu\nu}$ will then turn the observables into operators in quantum gravity.

We will explore two natural asymptotic limits. In one, we send the whole measurement apparatus to asymptotic infinity while keeping its size $L$ fixed. This will be relevant for memory correlators. In the other, we take the experiment to be asymptotically large so that it involves signalling between past and future infinity. This naturally captures Shapiro time delays.

It will be useful to change to notation better adapted to taking asymptotic limits and describing the celestial sphere. We will use the new notation in Sections \ref{sec:asymptotic_limit} and \ref{sec:memory_correlators}. For the convenience of the reader, we collect the main notational differences here:
\begin{align*}
&\text{\underline{Quantity}}&& \text{\underline{Bulk}}  && \text{\underline{Celestial}}
\\
&\text{Worldline separation}&& n^\mu && w^\mu
\\
&\text{Radial direction} && \mathbf{\hat{r}} &&\mathbf{\hat{n}}
\\
&\text{Angular direction} &&(\theta,\phi) && z^\mu = (1,\mathbf{\hat{n}})
\\
&\text{Displacements} && x_2-x_1 && z_2^A-z_1^A
\\
&\text{Time} && t  && \alpha=2u
\end{align*}
The main change is that on the celestial side, the null vector $z^\mu$ pierces the two-dimensional celestial sphere at location $z^A$ on the sphere, where $z^A$ are coordinates on the sphere, and $\mathbf{\hat n}^i(z^A)$ is the three-dimensional spatial unit vector specifying this same point. The arm separation denoted by $n$ in the finite-distance case becomes the spacelike Lorentz four-vector $w^\mu = (0,\mathbf{\hat w})$, or $w^A$ on the sphere. Greek and capital letter indices are exchanged with a projective vielbein.

\subsection{Sending the apparatus to infinity}

Displacement memory is often stated as a permanent change in a distance between two masses  \cite{Christodoulou:1991cr, Strominger:2017zoo}. The experimental measurement is typically described as some detector that measures distance or proper time and resides far from the gravitational wave source. However, the observables are usually stated as curvature tensors or specific components of the metric, sometimes in Bondi coordinates, rather than the explicit distance or time measures. While memory is a classical effect, our goal will be to understand how this connection between bulk and boundary observables is modified in quantum gravity. In this section, we will first give the connection in classical gravity.

We will start by making contact with the classic formulation of memory defined at asymptotic infinity. Having computed $\sigma$ in an arbitrary gauge, we are free to pick Bondi coordinates which are well-adapted to asymptotic quantities. We can then study the limit in which the detector is far from the origin, approaching the conformal boundary of spacetime, which is implemented by a large-radius limit. This will allow us to identify which asymptotic degree of freedom present in the Bondi metric encodes $\sigma$, and how it relates to the usual formulations of memory. This also strengthens the experimentally-relevant backstory for quantized memory operators. To model the detection of gravitational waves, the source is placed at the origin so that the faraway interferometer occupies a small, ideally infinitesimal, angle at large radius.

The metric in Bondi is written in terms of coordinates $u,r,z^A$ with $u=t-r$,
\begin{equation}
ds^2 = -du^2 - 2 du dr + r^2  \gamma_{AB} dz^A dz^B + r C_{AB}(u,z) dz^A dz^B+\cdots,
\end{equation}
where $\gamma_{AB}$ is the metric on the celestial sphere and $C_{AB}$ is the Bondi shear. The standard asymptotic limit is large $r$. Concretely, we will place the detector at $r=R$ and take $R$ large compared to $L$. As we consider translation-invariant setups, the origin is an unphysical coordinate location rather than a special point.

\begin{figure}[!ht]
\centering
\begin{tikzpicture}[scale=1.2, >=stealth, every node/.style={font=\small}]

\definecolor{worldline}{RGB}{31,86,137}
\definecolor{connector}{RGB}{190,82,30}
\definecolor{boundary}{RGB}{100,108,118}
\definecolor{nullray}{RGB}{28,128,64}

\coordinate (P) at (0.90,-0.26);
\coordinate (Q) at (1.63, -0.17);

%Boundary
%\draw(-3.30,0)--(0,3.20) node[above] {$i^+$}--(3.30,0);
%\draw(-3.30,0)--(0,-3.20) node [below] {$i^-$}--(3.30,0);
%\draw[dashed](3.30,0) arc (0:180:3.30 and 0.55);
%\draw (-3.30,0) arc (180:360:3.30 and 0.55);
%\node[right] at (3.30,0) {$i^0$};
\pgfmathsetmacro{\bdTheta}{asin(.55/3.20)}% tangency angle of the cone silhouette
\draw[rounded corners=3pt] (0,3.20) node[above] {$i^+$} -- (-3.30,0) -- (0,-3.20) node[below] {$i^-$};
\draw[rounded corners=3pt] (0,3.20) -- (3.30,0) -- (0,-3.20);
\draw[dashed](3.26,0) arc (0:180:3.26 and 0.55);
\draw (-3.26,0) arc (180:360:3.26 and 0.55);
\node[right] at (3.26,0) {$i^0$};
\node at (-1.90,1.80) {$\mathcal I^+$};
\node at (-1.90,-1.80) {$\mathcal I^-$};

%Inner circle
\draw[dashed](1.80,0) arc (0:180:1.80 and 0.30);
\draw        (-1.80,0) arc (180:360:1.80 and 0.30);
\node at (-1.2,0.05) {$r=R$};

%z line
\draw[nullray,thick,dashed](1.65,-0.48)--(0,3.20) node[pos=0.45,right]{$z$};

%worldlines
\draw[worldline,thick]
(0,-3.20).. controls (0,-2.50) and (0.90,-1.30) .. (P)
.. controls (0.90, 1.40) and (0, 2.50) .. (0,3.20);

\draw[worldline,thick]
(0,-3.20).. controls (0,-2.50) and (1.65,-1.30) .. (Q)
.. controls (1.65, 1.40) and (0, 2.50) .. (0,3.20);

\node[left] at (0.60,1.00) {$x_1(\tau)$};
\node[left] at (2.50,0.80) {$x_2(\tau)$};

%Radius and arm
\draw[->] (0,0)--(P) node[midway,above] {$R\,\mathbf{\hat n}$};
\draw[connector,very thick,->] (P)--(Q) node[midway, above=1pt, font=\footnotesize]{$L\,w^\mu$};

%Points
\fill (0,0) circle (1pt);
\fill (P) circle (1.5pt);
\fill (Q) circle (1.5pt);

\end{tikzpicture}

\caption{A diagram of the distance observable defined by two timelike worldlines (blue) $x_1, x_2$ at finite distance $r=R$ from the origin, embedded in the Penrose diagram of Minkowski spacetime. The first worldline is at an angle $\mathbf{\hat n}$ on the celestial sphere. The detector arm (orange) is of length $L$ along the direction $w^\mu$. The second worldline is at the same angular location $\mathbf{\hat n}$ up to $1/R$ corrections. Sending $R\rightarrow \infty$ keeping $L$ and $w$ fixed sends the detector apparatus to the conformal null boundary at the angle $\mathbf{\hat n}$ corresponding to the null ray $z=(1,\mathbf{\hat n})$ (green). The limiting detector is integrated along advanced and retarded time on $\scri^{\pm}$, along the null ray $z$, with a tangential polarization $w$. In our computations, we consider trivial past boundary conditions, and consider the effect of bulk fluctuations on the detector living along future null infinity $\scri^+$, similar to energy correlators in QFT.
}
\label{fig:penrose-detector}
\end{figure}
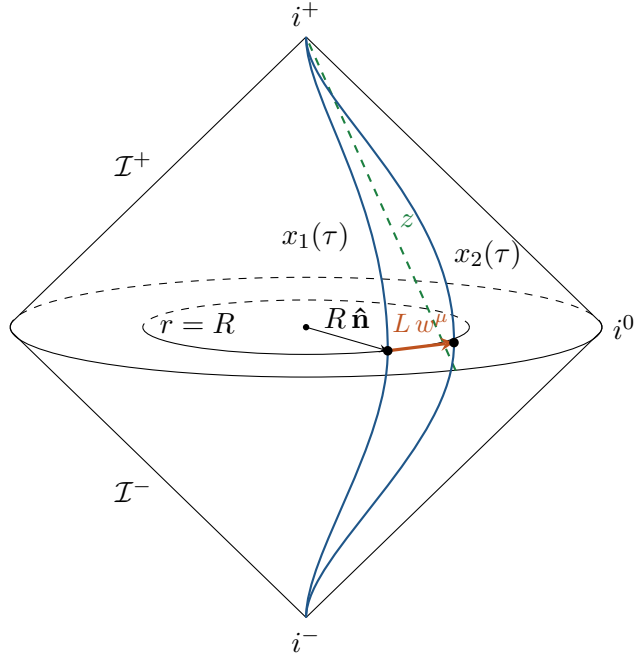

It will be useful to switch to the following notation to describe the asymptotic limit. We will need to parametrize both the location of the detector on the celestial sphere and its orientation. We will take the detector velocity, formerly $\dot{x}^{(0),\mu}$, to be the tangent vector $T^\mu$. The worldline separation, $n^\mu$ elsewhere, will be denoted by $w^\mu$. We send the detector at $r=R$ to infinity along a specific angular direction, which we parametrize by $z^\mu = z^\mu(z^A) = (1,\mathbf{\hat{n}}(z^A))$, where $z^\mu$ is a future-oriented null vector and $\mathbf{\hat{n}}^i$ is a three-dimensional unit vector parametrizing points on the celestial two-sphere $S^2$. The $z^A$ are the usual parametrization of the celestial sphere in Bondi coordinates, and we hope the overloaded notation with $z^\mu$ the null vector does not cause confusion. The tangent space of the celestial sphere is spanned by the two-dimensional vielbein $e^\mu_A = \partial_A z^\mu$, which comes with a projector onto the two-dimensional celestial sphere. The dual $e^A_\mu$ exchanges Cartesian displacements $\Delta x^\mu$ for angular displacements $\Delta z^A$ as $\Delta z^A = R^{-1} e^A_\mu \Delta x^\mu$ when $R$ is large. The detector orientation $w^\mu$ can lie tangent to the celestial sphere, $w \cdot z  = 0$, or orthogonal, $w \cdot z = 1$, or in between, $0< w \cdot z < 1$. For simplicity, we will restrict ourselves to the tangential case, and by definition, the orthogonal components give contributions that are subleading in $L/R$, which can also be accounted for in our approach. The orthogonal and mixed arm orientation cases give interesting connections to other memory components (the Bondi mass aspect and the divergence of the displacement memory), which we hope to elaborate on in the future.

In short, we will consider a detector located at null time $u$, radial location, $r=R$, and location $\mathbf{\hat{n}}$ on the celestial sphere, and oriented in direction $w$ tangent to the celestial sphere. The spacetime indices are $\mu = 0, \ldots ,3$, the Cartesian spatial indices are $ i = 1, \ldots, 3$, and the celestial sphere indices are $ A = 1,2$. The two worldlines are then specified as $\dot{x}_1^{(0),\mu} = \dot{x}_2^{(0),\mu} = T^\mu$. Holding these worldlines at fixed $r=R$ means that
\begin{align} 
x_1^{(0),\mu}(u) &= u T^\mu + R z^\mu  = (u+R, R\, \mathbf{\hat{n}}),
\\
x_2^{(0),\mu}(u) &= x_1^{(0),\mu}(u) + L w^\mu = (u+R, R\, \mathbf{\hat{n}}+L \mathbf{\hat {w}}).
\end{align}

Next, we will then send $R$ to infinity. The strict $R\rightarrow \infty$ limit (i.e. to order $(L/R)^0$) sends the entire detector to a single angle $z^\mu = (1,\mathbf{\hat{n}})$ on the celestial sphere. More generally, the detector worldlines converge to a single null line along the conformal boundary of spacetime. The setup and the limit are depicted in Figure~\ref{fig:penrose-detector}. 
We are interested in the component of the detector worldline that lies along future null infinity $\scri^+$, as the part that lies in the past does not receive a signal from fluctuations in the bulk. This null worldline is parametrized via $(u,z)\in \scri^+$. However, the leading correction in $L/R$ gives a small angular separation on the celestial sphere along the direction $w^\mu$. To leading order in $L/R$, the unperturbed spacelike geodesic along the detector arm connecting the two worldlines is given in Bondi coordinates as
\begin{equation}
x^{(0),\mu}_R(u,\lambda) = (u,R,z^A+\lambda L w^A /R),
\end{equation}
where $0 \leq \lambda \leq 1$ parametrizes the detector arm.

The proper distance $\sigma$ takes an especially simple form to leading order in $L/R$ under the condition that there are no contributions from metric perturbations of arbitrarily high frequency, which allows us to replace the integral over the detector arm by a simple average. Then in synchronous gauge, for instance,
\begin{equation}
\sigma(u;z,w) \approx L + \kappa \frac{L}{2}h_{ww}(r= R, u,z).
\end{equation}
In principle, the contributions from UV modes and any associated divergences must be treated separately. We estimate they will appear as corrections in $ \omega L^2/R$. But one can obtain a realistic detector by smearing over a finite-time window, in which case this approximation applies, as there are no longer UV divergences in the detector operator.

It is worth noting that metric perturbations can naively appear to invalidate this picture by displacing the detector from its fixed-$R$ location. However, since we work in perturbation theory, our formulas for $\sigma$ express such corrections naturally in terms of the unperturbed data, which in this case is a detector living at $r=R$. So we can blithely take the large-$R$ limit of our expressions.

In general, we can define the detector sitting at large radius $r=R$ as $\sigma_R(u;z,w)$. In principle, we can organize it into a large-$R$ expansion as 
\begin{equation}
\sigma_R(u;z,w)
=L+\sum_{n\geq1}\left(\frac{L}{R}\right)^{n}\s_{n}(u;z,w).
\label{eq:sigma R expansion}
\end{equation} There are two expansions that should not be conflated: the order of metric perturbations, and separately the order in the $L/R$ expansion. We are mainly interested in the leading order $L/R$ term which we will compute below and connect to the asymptotic shear. The subleading coefficients in $L/R$ are more challenging, depending on the metric data as well as the details of the detector, and we leave them as a future challenge. It is interesting to ask whether logarithmic or other nonanalytic terms may appear in this $1/R$ expansion; for example it is known that certain logarithms may appear in the asymptotic metric data~\cite{Winicour:1985pi,Chrusciel:1993hx}, which might enter $\sigma$. 

Now, let's proceed with taking the large $R$ limit of $\sigma$ and computing the leading $L/R$ correction $\sigma_1$. Throughout we are working to linear order in metric perturbation, so we want to compute $\sigma_R = L + \frac{L}{R} \sigma_1^{(1)} +\cO (1/R^2)$. To do this, we will need to compute the ingredients such as the Riemann tensor to linear order in metric perturbations and $1/R$. 
Using that
\begin{equation}
R_{0w0w}^{(1)} = -\frac{1}{2R}\partial_u^2 C_{ww} +\cO(1/R^2)
\end{equation}
to leading order in $1/R$, with $\partial_u = \partial_T$ at fixed $r$, we then have
\begin{equation}
\sigma_R(u;z,w) = L+\frac{L}{2R} \int_{-\infty}^u du'(u-u')\int_0^1 d\lambda \partial_{u'}^2 C_{ww}(u',z+\l \tfrac{L w}{R}) +\cO(1/R^2),
\end{equation}
or
\begin{equation}
\sigma_R(u;z,w) = L+\frac{L}{2R} \int_{-\infty}^u du'\int_0^1 d\lambda N_{ww}(u',z+\l \tfrac{L w}{R}) +\cO(1/R^2),
\end{equation}
upon integrating by parts, where $N_{AB}(u,z) = \partial_u C_{AB}(u,z)$ is the Bondi News tensor. The boundary term we dropped is $-\lim_{u_0 \rightarrow -\infty } (u-u_0) N_{ww}(u_0,z)$, which is zero as long as $N_{ww}$ decays faster than $1/|u_0|$ at early times. This is an assumption about the physical radiative data in addition to asymptotic flatness, which only constrains the falloff in $r$ and allows any $N_{AB}(u,z)$. We emphasize that this is not a choice of BMS frame, and is invariant under supertranslations.

Finally, in the small $L/R$ regime, the integral over $\lambda$ can be approximated by the value on the reference ray $z$,
\be
\int_0^1 d\lambda N_{ww}(u',z+\l \tfrac{L w}{R}) = N_{ww}(u',z) + \cO(L/R) \, .
\ee 
For high frequencies, there will be corrections in $\omega L^2/R$, as discussed above, which are already suppressed in $L/R$.\footnote{
More precisely, if we keep track of the radial displacement of the arm $r(\l) = \sqrt{R^2+L^2\l^2} \approx R + \l^2 L^2/2R$, there is a resulting retarded time displacement $u(\l) = t-r(\l) \approx u - \l^2 L^2/2R$. Plugging this into the integral over the arm geodesic, we have
$$\int_0^1 d\lambda\,N_{ww}(u(\l),z(\l))=N_{ww}(u,z)+\frac{L}{2R}\,w^A D_A N_{ww}(u,z)-\frac{L^2}{6R}\,\partial_{u}N_{ww}(u,z)+\cO(L^2/R^2).$$ 
From this we can read off the leading $L/R$ and $\omega L^2/R$ corrections to the arm response, as claimed. Note also that corrections from high frequency modes $N_{ww}(u,z)\sim e^{i\omega u} N_{ww}(z)$ with $\omega L^2/R \gg 1$ can also be evaluated via a saddle-point as $\int_0^1 d\lambda\, e^{i\omega ( \l^2 L^2/2R)} N_{ww}(z) \sim \cO(\sqrt{R/\omega L^2}) $ and are suppressed.
} With this approximation, we simply have
\begin{align}
\sigma_R(u;z,w) &= L+\frac{L}{2R} \int_{-\infty}^u du'(u-u') \partial_{u'}^2 C_{ww}(u',z) +\cO(1/R^2),
\\
&= L+\frac{L}{2R} \int_{-\infty}^u du' N_{ww}(u',z) +\cO(1/R^2).
\end{align}
Performing the $u'$ integral gives
\begin{equation}
\sigma_R(u;z,w) = L+\frac{L}{2R}  \left[ C_{ww}(u,z) - C_{ww}(-\infty,z)\right]+\cO(1/R^2).
\end{equation}
The boundary condition for the initial shear $C(-\infty,z)$ is of a different nature than the News falloff discussed above. When the News vanishes at $u\to -\infty$, the initial shear is independent of $u$, and when magnetic-parity shear is also absent at early times, it can be removed by a supertranslation. Therefore, we pick the BMS frame $C(-\infty,z)=0$, so that 
\begin{equation}
\sigma_R(u;z,w) = L+\frac{L}{2R}   C_{ww}(u,z) +\cO(1/R^2).
\end{equation} Unlike the News falloff, this is a choice of BMS frame and not a condition on the radiative data. The difference $\left[ C_{ww}(u,z) - C_{ww}(-\infty,z)\right]$ is BMS frame independent.  

In terms of the metric, the asymptotic form of the $\mathcal{O}(\kappa)$ correction  $\sigma_1^{(1)}$ is 
\begin{equation}
\sigma_{1}^{(1)}(u;z,w) = \lim_{R\rightarrow \infty} \frac{R}{L} (\sigma_R(u;z,w)-L)  = \frac{1}{2} C_{ww}(u,z) = \frac{\kappa}{2} h_{ww}(2u,z),
\end{equation}
where $h$ here is specifically the radiative piece of $h$, denoted $h^\perp$ below, to which we restrict ourselves in this section. We will explain how to extract $h^\perp$ later, see Eq.~\eqref{eq:asymptotic metric field}.

We can also write differences of distances, for instance
\begin{equation}
\sigma_R(u;z,w)-\sigma_R(u';z,w) = \frac{L}{2R}(C_{ww}(u,z) - C_{ww}(u',z)) + \cO(1/R^2,\kappa^2).
\end{equation}
This change in distance is the finite-time version of the classic memory observable. As expected, this difference is independent of the BMS data $C(-\infty,z)$ we set to zero earlier. Sending $u \rightarrow \infty, u' \rightarrow -\infty$, this recovers the standard classical memory due to a passing gravitational wave,
\begin{equation}
\sigma_R(\infty;z,w)-\sigma_R(-\infty;z,w) \approx \frac{L}{2R}(C_{ww}(\infty) - C_{ww}(-\infty)) 
=
\frac{L}{2R} \int_{-\infty}^\infty du \, \partial_u C_{ww}
=
\frac{L}{2R} \int du N_{ww}.
\end{equation}

In essence, the procedure we carried out amounts to taking the large-distance limit by stripping off precisely the decaying power of $R$ to obtain a finite quantity. We will unpack this general mechanism in  Section \ref{sec:detectors}.

We end with a related comment. We have just taken the asymptotic limit of $\sigma$ in Bondi coordinates. This is natural when the perturbation $h$ has a compact source residing near $r=0$. In the quantum gravity case, we can have non-zero fluctuations with no sources, so the $r=0$ point is not physically special. We will now consider a different asymptotic limit that captures such distance fluctuations in a way that may more closely resembles a realistic experimental protocol.

Consider two distance measurements occurring at times $\tau_1, \tau_2$ with $\tau_2-\tau_1 \gg L$. One can  form an observable $\Delta \sigma$ that captures the change in the separation between worldlines from time $\tau_1$ to time $\tau_2$, which is $\Delta \sigma(\tau_2, \tau_1) \equiv \sigma(\tau_2)- \sigma(\tau_1)$. This observable captures displacement memory in classical gravity. We can now ask whether this quantity fluctuates in the quantum gravity vacuum. Using translation invariance to send $\tau_1 \rightarrow 0$ and $\tau_2 \rightarrow \tau_2-\tau_1 \equiv \delta \tau$, the fluctuations of $\Delta \sigma$ are captured by $\braket{\Delta \sigma(\delta \tau +\tau,\tau) \Delta \sigma(\delta \tau,0) }_c$. This describes the correlation between two change-of-distance measurements that are initiated a $\tau \ll \delta \tau$ proper time apart, where each measurement is the change in distance over a time interval $\delta \tau$. The measurements coincide at $\tau = 0$.

We can now compute the relevant late-time correlator of $\Delta \sigma$. As we computed earlier, $\braket{\sigma(t)\sigma(0)}_c \sim -(\kappa^{2} L^2)/(30\pi^{2}t^{2})$ for $t \gg L$ in synchronous coordinates. This and the time-translation invariance of the correlator allow us to easily determine the $\delta \tau \gg \tau$ limit,
\begin{equation}
\delta \tau \gg \tau: ~~~~~~~\braket{\Delta \sigma(\delta \tau +\tau,\tau) \Delta \sigma(\delta \tau,0) }_c 
\approx 
2 \braket{\sigma(\tau)\sigma(0)}_c,
\end{equation}
as the cross terms decay at large $\delta \tau$ because the vacuum correlations decay with separation. The commutator is similarly nonzero, confirming the quantum nature of this observable. One can regard this as a quantum extension of gravitational memory that is induced by vacuum fluctuations, although the phenomenon being captured is entirely different from the classical case. The effect here comes entirely from short-range correlations of distance rather than the low-frequency effects that yield classical memory.

As explained earlier, it is more realistic to measure round trip time than proper distance, and the two coincide for low-frequency signals. The round-trip time observable analogous to $\Delta \sigma$ is $\Delta \tau(\delta \tau_1,0) \equiv \tau(\delta \tau_1)-\tau(0)$ in the large $\delta \tau_1$ limit, where we recall that $\tau_1$ is the start time of the round-trip protocol, which has elapsed time $\tau \equiv \tau_3-\tau_1$. The vacuum correlation is
\begin{equation}
\delta \tau_1 \gg \tau_1: ~~~~~~~ \braket{\Delta \tau(\delta \tau_1+\tau_1,\tau_1) \Delta \tau(\delta \tau_1,0)}_c 
\approx 2 \braket{\tau(\tau_1)\tau(0)}_c.
\end{equation}
See \cite{Carney:2024wnp, Sivaramakrishnan:2025srr} for explicit expressions. Once again, this quantity and the commutator are both non-zero. No IR divergences appear, as expected for a finite-distance measurement.

\subsection{Making the apparatus infinitely large}
\label{sec:detectors}

We will shortly find ourselves taking asymptotic limits that push operators to infinity inside correlation functions. In order to extract a finite result, we will multiply by the asymptotic parameter to the appropriate power to cancel the large-distance decay. In doing this, we must be careful with the coincident point and null singularities. To understand the basic mechanics of this limit, it suffices to focus on a free scalar two-point function, where the scalar can be thought of as an instantaneous detector. We can take both operators to future infinity, or one to future and the other to past infinity. The second option corresponds to taking the experimental setup, which is comprised of the detectors and the source of the signal they measure, to be infinitely large so that it can be described via asymptotic detectors. 

In this section we identify some simple observables at asymptotic infinity that encode the arrival time of null signals. We will focus on single null rays in flat space, but we expect these ideas to generalize to weakly-curved asymptotically flat space and multiple null rays. The basic idea is that if certain correlators have distinctive features, such as a singularity or a sharp change, the arrival time of a signal can be extracted from these correlators. 

Consider sending quanta of a massless scalar field $\phi$ from past null infinity to future null infinity, $\scri^-$ and $\scri^+$. We take $\phi$ to be a free field, and we will consider the effects of gravitational fluctuations to such observables by returning to the EFT later on. A simple correlator that is sensitive to the arrival of the signal is the two-point function of the field $\phi$, inserted at $\scri^-$ and $\scri^+$. Defining
\begin{equation}
\label{eq:asymptotic scalar field}
    \phi(\alpha,z)= \lim_{R\to \infty} R^{\Delta_\phi} \phi(x+R z),
\end{equation} where $\a=-2x\cdot z$ and $z$ is a future-pointing null vector, we have that $\phi(\alpha,\pm z)$ is a local operator inserted at $\scri^{\pm}$. The coordinate $\a$ is the retarded / advanced time at $\scri^{\pm} \cong \R \times S^{d-2}$, and $z$ parametrizes a direction on the celestial sphere $S^{d-2}$ via $z=(1,\mathbf{\hat{n}})$ in coordinates $(x^0,\vec x)$. Now, let's consider a correlator in free theory (or any CFT) of one field inserted at $\scri^-$ and one at $\scri^+$,\footnote{The phase can be understood as arising from translating one of the operators by one Minkowski patch on the Lorentzian cylinder, i.e. the action of the operator $\cT$~\cite{Kravchuk:2018htv}.}
\begin{equation}
\<\f(\a_1,z_1) \f(\a_2,-z_2)\> = e^{-i \pi \De_\f}\frac{N_\f}{(-2z_1\cdot z_2)^{\De_\f}}\,.
\end{equation}
The normalization can be chosen as $N_\f=\frac{\Gamma (\Delta_\phi )}{\pi ^{d/2} 2^{d-2 \Delta_\phi }}$ to correspond to the standard momentum space propagator $i/p^2$~\cite{Caron-Huot:2022eqs}. 
We can choose to parametrize the celestial sphere using Poincare coordinates, via $z=(z^+,z^-,\vec z)=(1,y^2,\vec y)$, where $\vec y$ can be thought of as transverse coordinates in the bulk.\footnote{This is the $z^+ = 1$ section of the null cone. Elsewhere, we use the section $z^0 = 1$, with $z^\mu = (1,\mathbf{\hat{n}})$. The two sections differ by a $z$-dependent rescaling of $z^\mu$, and therefore of $\alpha$, and so the difference does not affect conformally-invariant statements in this subsection.}
\begin{equation}
\<\f(\a_1,z_1) \f(\a_2,-z_2)\> = e^{-i \pi \De_\f}\frac{N_\f}{[(\vec y_1 - \vec y_2)^2]^{\De_\f}}\, .
\end{equation} These expressions are for generic kinematics and we have not been careful about $i\varepsilon$ prescriptions or distributional terms. We will now analyze these pieces directly by using the explicit mode expansion. We will encounter a delta function $\delta^{(d-2)}(z_1,z_2)$ that captures support at coincident angles, which depends on $\alpha_{12}$. By contrast, the term above arises from soft modes and ends up being removed by an infrared cutoff. The two expressions will agree once we remove the zero modes by taking derivatives.

To simplify the problem, we can use conformal invariance to map the system to the two point function restricted to a null plane through the origin. Free theory being conformal naturally lives on the Lorentzian cylinder, so the field approaching $\scri^-$ can be translated via the operator $\cT$~\cite{Kravchuk:2018htv} to the next Poincare patch, at the cost of the phase $e^{-i \pi \De_\f}$ we have seen above. In the new configuration, we have two scalars approaching the null plane $\scri^+$, one from the future and one from the past. Now, we can perform a conformal transformation from the original coordinates $x^\mu$ to new coordinates $y^\mu(x)$ that maps $\scri^+$ to the null plane through the origin given by $y^+=0$ (see~\cite{Hofman:2008ar}). The object we need to analyze is simply the two point function on the null plane $y^+=0$, where we have the identification $y^- = \a$, and transverse coordinates are $\vec y$, since 
\be
\<0|\f(\a_1,z_1) \f(\a_2,-z_2)|0\>
 = e^{-i \pi \De_\f} \<0|\f(y_1) \f(y_2)|0\> \Big|_{y_i^+=0, \ y_i^- = \a_i}.
\ee
Being careful with the $i\varepsilon$ prescription, the Wightman function is given by
\be
\<0|\f(y_1^+=0,y_1^-,\vec y_1) \f(y_2^+=0,y_2^-, \vec y_2)|0\> 
 = \frac{N_\phi}{\left[{\vec y}_{12}^2 + i \varepsilon y_{12}^- + \varepsilon^2\right]^{\De_\f}}\, .
\ee  When ${\vec y}_{12}\ne0$, we see that the answer is $N_\phi/{\vec y}_{12}^{~2\De_\f}$. However, there are distributional terms that we can read from the $i\varepsilon$ prescription. In fact, there is a singularity when $\vec y_{12} = 0$. To get a nice distribution, we can take a derivative with respect to $\a_1=y_1^-$. Let's set $d=4$ and $\De_\f=1$. We have
\be
&\<0| \partial_{-} \f(y_1^+=0,y_1^-,\vec y_1) \f(y_2^+=0,y_2^-, \vec y_2)|0\> 
 = \frac{-i\varepsilon N_\phi}{\left[{\vec y}_{12}^2 + i \varepsilon y_{12}^- + \varepsilon^2\right]^{2}}\, .
\ee Now, this distribution is zero when ${\vec y}_{12} \neq 0$, and singular otherwise. To understand the distributional term, we integrate in ${\vec y}$ against a test function that only has support at ${\vec y}_{12} = 0$. Using $\int d^2 y (\vec{y}^2+a)^{-2} = \pi/a$ with $a = \varepsilon^2 +  i\varepsilon y_{12}^-$ and $N_\phi = 1/(4\pi^2)$, we have
\be
 \<0| \partial_{-} \f(y_1^+=0,y_1^-,\vec y_1) \f(y_2^+=0,y_2^-, \vec y_2)|0\> 
& = - \pi N_\phi \delta^{(2)}(\vec y_{12}) \frac{1}{2} \frac{1}{ y_{12}^- - i \varepsilon}
\nn \\ &  = - \frac{1}{4\pi} \delta^{(2)}(\vec y_{12}) \frac{1}{2} \left( \cP\left[\frac{1}{y_{12}^-}\right] + i \pi \de( y_{12}^-) \right),
\label{eq:phi two point arrival signal}
\ee where $\cP$ is the Cauchy principal value distribution.

Let's compute this again in free theory using the mode expansion of the field $\phi$. One computes
\be
\f(\a,z) = \frac{1}{2(2\pi)^{d/2}}\int_0^\infty d\beta \beta^{\frac{d-4}{2}} &\left(i e^{-i\frac{\pi d}{4}} e^{-i\frac{\a\b}{2}} a(\b z) -i e^{i\frac{\pi d}{4}} e^{i\frac{\a\b}{2}} a^\dagger(\b z) \right),
\ee
and similarly
\be
\f(\a,-z) = \frac{1}{2(2\pi)^{d/2}}\int_0^\infty d\beta \beta^{\frac{d-4}{2}} &\left(-i e^{i\frac{\pi d}{4}} e^{-i\frac{\a\b}{2}} a(\b z) +i e^{-i\frac{\pi d}{4}} e^{i\frac{\a\b}{2}} a^\dagger(\b z) \right).
\ee
We have used a relativistic normalization of the modes, with 
\begin{equation}
[a(\b_1 z_1), a^\dagger (\b_2 z_2)] = 2 \b_1^{3-d} (2\pi)^{d-1}\delta(\b_1-\b_2) \de^{d-2}_{S^{d-2}}(z_1,z_2).
\end{equation}
For simplicity, we define the following infrared-regulated kernel:
\begin{align}
\mathcal{I}_0(s) \equiv \lim_{\varepsilon \rightarrow 0^+} \int_\Lambda^\infty \frac{d\omega}{\omega} e^{-i\omega(s-i \varepsilon )} =& -\gamma-\log(i\Lambda s) + \mathcal{O}(\Lambda),
\nonumber
\\
\mathcal{I}_0'(s) = -\frac{1}{s-i\varepsilon}, ~~~~~~ \mathcal{I}_0''(s) = \frac{1}{(s-i\varepsilon)^2}, 
\end{align}
with the IR regulator $\Lambda > 0$, and where in the bottom line we have safely sent $\Lambda \rightarrow 0$. Shifting $\Lambda$ by a constant is captured by the soft mode.
Sending $\alpha_1 \rightarrow \alpha_1 - i\varepsilon/2$ with $\varepsilon > 0$, which defines the Wightman function, the correlator with an IR regulator $\Lambda > 0$ becomes 
\be
\<0| \f(\a_1,z_1)\f(\a_2,-z_2)|0\> 
&= -\frac{e^{-i \frac{\pi d}{2}}}{4\pi} \de^{(d-2)}_{S^{d-2}}(z_1,z_2) \int_\Lambda^\infty \frac{d\b}{\b} e^{\frac{i}{2}\b(\a_2-\a_1) - \varepsilon \beta}
\nn \\ 
&= - \frac{e^{-i \frac{\pi d}{2}}}{4\pi}\delta^{(d-2)}_{S^{d-2}}(z_1,z_2) \mathcal{I}_0\left( \frac{\alpha_1-\alpha_2}{2} \right)
\nn
\\&  = \frac{e^{-i \frac{\pi d}{2}}}{4\pi} \de^{(d-2)}_{S^{d-2}}(z_1,z_2)\left( \gamma + \log\left(\frac{i \Lambda(\a_1-\a_2)}{2} \right)\right)
\, 
\label{asymptotic_scalar_2pt_wightman}
\ee
to leading order in $\Lambda$ and in the $\varepsilon \rightarrow 0$ limit.

There is an IR divergence arising from the small-$\beta$ region of the integrand, but if we consider instead an insertion of $\partial_\alpha \phi$ rather than $\phi$, the IR divergence is not present. Concretely, we have
\be
\<0| \partial_{\alpha_1}\f(\a_1,z_1)\f(\a_2,-z_2)|0\> 
&  = -\frac{ e^{-i \frac{\pi d}{2}}}{8\pi} \de^{(d-2)}_{S^{d-2}}(z_1,z_2) \mathcal{I}_0'\left(\frac{\alpha_1-\alpha_2}{2} \right)
 \nn \\ & = -\frac{ e^{-i \frac{\pi d}{2}}}{4\pi} \de^{(d-2)}_{S^{d-2}}(z_1,z_2) \frac{1}{\a_2-\a_1 + i \e}
\nn \\ &  = -\frac{ e^{-i \frac{\pi d}{2}}}{4\pi} \de^{(d-2)}_{S^{d-2}}(z_1,z_2)
\left( \cP\left[\frac{1}{\a_2-\a_1} \right] -  i \pi \delta(\a_2-\a_1) \right),
\ee in precise agreement with~\eqref{eq:phi two point arrival signal}. We can think of this observable as roughly sourcing a quantum $\f(\a_2,-z_2)$ at $\scri^-$, and then having a local detector $\partial_{\alpha_1}\f(\a_1,z_1)$ that is sensitive to the jump when the signal arrives at $\scri^+$. The signal is detected when the retarded time $\a_1$ of the detector hits the advanced time $\a_2$ of the source, corresponding to the jump across $\de(\a_2-\a_1)$, and is only there precisely when the angles coincide $z_1=z_2$ (suitably smeared for a wavepacket).

Finally, note also that another interesting observable is the correlation function with one operator at infinity and the other at a finite point,
\be
\<\f(\a,z)\f(x')\> = \lim_{R\to\infty} R^{\De_\f} \frac{1}{\left( x +R z - x'\right)^{2\De_\f}} = \frac{1}{\left[2z\cdot(x-x')\right]^{\De_\f}}
= \frac{1}{[\a' -\a]^{\De_\f}},
\ee where $\a=-2z\cdot x$ as before and $\a' = -2z \cdot x'$ is the retarded time of the signal from point $x'$. Once again, this exhibits a jump (singularity) when the retarded times coincide.

Similarly, we can study the two point function of asymptotic metric fields. The logic is identical, with the only difference being the polarizations. We give the relevant formulae here. In de Donder gauge, the Wightman function is \cite{Donoghue:1995cz}
\begin{equation}
\braket{0|h_{\mu\nu}(t,\mathbf{x}) h_{\rho\sigma}(0)|0} = P_{\mu\nu,\rho\sigma} \frac{1}{4\pi^2} \frac{1}{\mathbf{x}^2-(t-i\varepsilon)^2},~~~~~ P_{\mu\nu,\rho\sigma} = \frac{1}{2}(\eta_{\mu \rho} \eta_{\nu\sigma} + \eta_{\mu\sigma} \eta_{\nu \rho}-\eta_{\mu\nu} \eta_{\rho\sigma}),
\end{equation}
where we use the normalization
\begin{align}
h_{\mu\nu}(x) &= \sum_s \int \frac{d^3 k}{(2\pi)^3 2\omega}\left( \epsilon^s_{\mu\nu}(k) a_s(k) e^{i k \cdot x} + \epsilon^{s*}_{\mu\nu} a^\dagger_s(k) e^{-ik \cdot x} \right),
\\
[a_s(k),a^\dagger_{s'}(k')] &= 2\omega (2\pi)^3 \delta^{(3)}(\mathbf{k} - \mathbf{k}') \delta_{ss'}.
\end{align}
We now want to define the asymptotic metric field, analogous to~\eqref{eq:asymptotic scalar field}, which gives\footnote{Note here that $\Delta_h = \frac{d-2}{2} = 1$ for $d=4$.}
\begin{equation}
h^{{\rm raw}}_{\mu\nu}(\alpha,\pm z) =\lim_{R\to\infty}R\, h_{\mu\nu}(x\pm Rz),\qquad h^{\perp}_{\mu\nu}(\alpha,\pm z) =\lim_{R\to\infty}
R\, \Pi^\perp_{\mu\nu}{}^{\rho\sigma}(z)h_{\rho\sigma}(x\pm Rz).
\label{eq:asymptotic metric field}
\end{equation}
The field $h^{\rm raw}$ is the naive asymptotic limit, which is gauge dependent: it is not diffeomorphism invariant and in particular transforms under the residual small diffeomorphisms in de Donder gauge. The projected field $h^\perp$ is invariant under small diffeomorphisms that preserve the $1/R$ falloff, thanks to the projector 
$\Pi^\perp_{\mu\nu,\rho\s}(z) = \sum_s \epsilon^s_{\mu\nu}(z) \epsilon^{s*}_{\rho\s}(z)$,\footnote{
The projector implements the projection onto the usual physical propagating polarizations of the graviton and can be written geometrically as
\be
\Pi^\perp_{\mu\nu,\rho\sigma}(z)
&=\frac12\left(q_{\mu\rho}q_{\nu\sigma}
+q_{\mu\sigma}q_{\nu\rho}
-q_{\mu\nu}q_{\rho\sigma}\right),
\ee where $q_{\mu\nu}
=\eta_{\mu\nu}
+\frac12\left(z_\mu\bar z_\nu+\bar z_\mu z_\nu\right)$ is the screen projector with the reference null vector $\bar z^\mu = 2T^\mu-z^\mu$.
}
and corresponds to the radiative degrees of freedom.\footnote{
For the tangential arm setup we are considering, with $w\cdot z=0$, the de Donder gauge metric automatically projects to the physical components to leading order in $1/R$, i.e. $h^{\rm raw}_{ww} = h^\perp_{ww}$. However, for more general arm orientations with components in the radial direction or to higher order in $1/R$, this does not hold and one needs to impose the physical projector $\Pi^\perp$.
}
In Bondi gauge, the projected field is precisely the shear,
\begin{equation}
C_{AB}(u,z)=\kappa e_A^\mu e_B^\nu h^{\perp}_{\mu\nu}(2u,z),\qquad \kappa h^{\perp}_{\mu\nu}(\alpha,z)=e^A_\mu e^B_\nu C_{AB}(\alpha/2,z).
\label{eq:Bondi matching}
\end{equation}

Using the same stationary-phase approximation as in the scalar case, the mode expansion of the asymptotic metric field is~\cite{Herrmann:2024yai}
\begin{equation}
h^\perp_{\mu\nu}(\alpha, \pm z) = \mp \frac{i}{8\pi^2} \sum_s \int_0^\infty d\omega
\left(
\epsilon^s_{\mu\nu}(z) e^{- i \omega \alpha/2} a_s(\omega z) - \epsilon^{s*}_{\mu\nu}(z) e^{i\omega \alpha/2} a^\dagger_s(\omega z)
\right),
\end{equation}
where $a_s$ is the mode with $\mathbf{k} = \omega \hat{\mathbf{n}}$, and the relative sign between the future and past null infinity is the phase $e^{-i \pi \Delta_h} = -1$. The commutator of these raising and lowering operators is $2(2\pi)^3 \omega_1^{-1}\delta(\omega_1-\omega_2)\delta^{(2)}_{S^2}(z_1,z_2) \delta_{ss'}$, in which case the asymptotic two-point function simplifies to $\mathcal{I}_0$, giving\footnote{In de Donder gauge, without imposing the projector $\Pi^\perp$, $\<h^{\rm raw}_{ww}h^{\rm raw}_{w'w'}\> \propto P_{ww,w'w'}$ instead, which happens to equal $\Pi^\perp_{ww,w'w'}$ for tangential arm $w$. This is no longer true with an arm with a radial component, and would yield an unphysical answer without the projector. Also, note that $\Pi^\perp\cdot P \cdot \Pi^\perp = \Pi^\perp$.} 
\begin{align}
\braket{0|h_{ww}^\perp(\alpha_1,z_1) h_{w'w'}^\perp(\alpha_2, \pm z_2)|0} &= \pm \frac{1}{4\pi} \Pi^\perp_{ww,w'w'} \delta^{(2)}_{S^2}(z_1,z_2) \mathcal{I}_0(\alpha_{12}/2) \nn
\\ &= \pm \frac{1}{4\pi} \left((w\cdot w')^2-\tfrac12 \right) \delta^{(2)}_{S^2}(z_1,z_2) \mathcal{I}_0(\alpha_{12}/2),
\label{hh_asymptotic_2pt}
\end{align}
where $\delta^{(2)}_{S^2}$ is the normalized sphere delta function. Note that we get the expected behavior for the graviton polarizations: the signal is maximized for overlapping polarizations (i.e. $\times \times$ or $+ +$) with $w\cdot w' =0, \pm 1$, and vanishes for orthogonal polarizations (i.e. $\times +$ or $+\times$) with $w\cdot w'=\pm \frac{1}{\sqrt{2}}$. Derivatives turn this two-point function into the same inverse powers as in the scalar case. These agree with the form obtained by using asymptotic quantization, and with the explicit expressions obtained in \cite{Liu:2023gwa}. Note that our formulas also encompass correlators with a field at $\scri^+$ and another at $\scri^-$.

%%%%

So far we have only discussed free two-point functions, whose singularities lie along null rays. When we perturbatively include gravitational effects, the light cone will deform, changing the location of the null singularity. To account for this effect, one must however start with diffeomorphism-invariant  operators  in the bulk. An example is the gravitationally-dressed operator $\phi(x(\tau))$ inserted at time $\tau$ along the worldline $x$. See \cite{Cheung:2026euf} for further details and the singularity structure. It would be interesting if Landau diagrams could be used to diagnose the location of the singularity without having to compute the full correlator.

\section{Geometric observables at asymptotic infinity}
\label{sec:asymptotic_observables}

We have analysed asymptotic observables that capture quantized extensions of memory and the arrival times of null signals. Now, we will return to the gravitational effective field theory and compute perturbative quantum gravity corrections to these observables, including correlators and matrix elements. For both memory and time delays, we will use the asymptotic form of the geometric observable we have defined, rather than a more realistic model of an asymptotic detector that more indirectly encodes such geometric observables. The geometric definitions allow us to directly compute quantum gravity extensions of memory and time delays without any detour through an auxiliary scattering or detector setup. The upshot is that we obtain memory and Shapiro time delays as independently-defined operators in perturbative quantum gravity.

We leave further strengthening the connection between asymptotic observables, including S-matrices and correlators, and time-delay type observables to future work. For work on detectors in quantum gravity at asymptotic infinity see \cite{Herrmann:2024yai, Caron-Huot:2023vxl, Caron-Huot:2025ymc} and at finite distance see \cite{Cheung:2026euf}. The process described here resembles the extrapolate dictionary in AdS/CFT, and it would be illuminating to see if a position-space procedure like this one does in fact produce the various known out-of-time-order asymptotic observables \cite{Caron-Huot:2023vxl, Caron-Huot:2025ymc}. For this, a path integral description may be more natural \cite{Kim:2023qbl}.

\subsection{Memory}
\label{sec:memory_correlators}

Now that we have obtained $\sigma$ in terms of $C_{AB}$, or the leading radiative asymptotic data of $h_{\mu\nu}$, we can obtain correlators and matrix elements of $\sigma$ from those of $C_{AB}$ or $h_{\mu\nu}$. Note that we have not gravitationally dressed the states, and it is unclear a priori if these correlators will end up being finite. We also reiterate that in what follows, we are only computing the radiative contribution to $\sigma$.

First, we will summarize the necessary ingredients. The vacuum two-point function of the graviton was given in \eqref{hh_asymptotic_2pt}. We will also need the matrix element 
\begin{equation}
\braket{0|h_{\mu\nu}(x)|h^s(k)} = P_{\mu\rho} P_{\nu\sigma}\epsilon_s^{\rho\sigma} e^{i k x},
\end{equation}
where we absorb numerical factors such as $2\pi$ into the definition of the state for convenience. Recall that in the construction of dressed operators we use, see Appendix \ref{app:geometric_observables_arbitrary_diffs_and_initializations}, these projectors enter when contracting with external gravitons in synchronous gauge, and render such matrix elements automatically diffeomorphism invariant.

We start from the matrix element of $\sigma_R$ in the small $L/R$ approximation, $\sigma_1$, although we will omit these subscripts in this section for convenience. We have shown this quantity is an insertion of $h_{ww}$,
\begin{equation}
\braket{0|\sigma(\tau)|h(k)}  = \frac{\kappa L}{2}\braket{0|h_{ww}(\tau)|h(k)} = \frac{\kappa L}{2} e^{i k x(\tau)}P_{w\rho} P_{w\sigma}\epsilon^{\rho\sigma}.
\end{equation}
We pass to Bondi coordinates,
\begin{equation}
x^\mu = (u+R, R \, \mathbf{\hat{r}}), ~~~ k^\mu = \omega(1, \mathbf{\hat{k}}),
\end{equation} 
so that $k \cdot x^{(0)} = -\omega(u + R(1 -  \mathbf{\hat{k}} \cdot \mathbf{\hat{r}}))$. The projector $P_{\mu\rho}$ is evaluated with $\dot{x}^{(0),\mu} = (1,0,0,0)$ and $k \cdot \dot{x}^{(0)} = -\omega$. Gauge invariance remains enforced automatically by $P_{\mu\rho}k^\rho = 0$, which holds in any coordinate system. The several approximations we have made in obtaining the asymptotic form of $\sigma^{(1)}$ have preserved this manifest diffeomorphism invariance.

To obtain a normalizable state, we must integrate over $k$. The standard stationary phase approximation at large $R$ we have used can be summarized succinctly as
\begin{equation}
e^{i k \cdot x} \rightarrow -2\pi i \frac{e^{-i \omega u} }{R\omega} \delta^{(2)}_{S^2}(\mathbf{\hat{k}}, \mathbf{\hat{r}}), 
\end{equation}
the delta function on the celestial sphere. At large $R$, we therefore have the finite quantity
\begin{equation}
\frac{2R}{L} \braket{0|\sigma^{(1)}(\tau)|h(k)} =
-\frac{2\pi i \kappa}{\omega} \delta^{(2)}_{S^2}(\mathbf{\hat{k}},\mathbf{\hat{r}})
e^{-i \omega u} P_{w\rho} P_{w \sigma} \epsilon^{\rho \sigma},
\label{CAB_h_matrix_element}
\end{equation}
which is nonvanishing at large $R$. The delta function means this is only non-zero when the spatial orientation of $k$ aligns with the angle $\mathbf{\hat{r}}$ at which we sent the detector to asymptotic infinity. Mechanically, this arises from contraction of $\sigma$ with the external on-shell $h$. When we send the detector to infinity, it does not detect the incoming $h$ unless the two are inserted at the same location at infinity. 

In order to determine whether the matrix element corresponds to a physical observable, we must first prepare a normalizable state 
\begin{equation}
\ket{\Phi} = \int d^3k \, \Phi(k)\ket{h(k)} = \int \omega^2 d\omega \int d \Omega_{2} \Phi(\omega, \mathbf{\hat{k}})\ket{h(k)}. 
\end{equation}
We then have
\begin{equation}
\frac{2R}{L} \braket{0|\sigma^{(1)}(\tau)|\Phi}
=
\braket{0|C_{ww}|\Phi}   = \kappa
\int d\omega \, \omega
\Phi(\omega,\mathbf{\hat{r}}) e^{-i \omega u} P_{w\mu}P_{w\nu}\epsilon^{\mu\nu},
\end{equation}
which no longer has the naive soft divergence.

Note that $\braket{0|\sigma^{(1)}(\tau)|\Phi}$ decays as $1/R$. We took the large $R$ limit but even though $r=0$ was not a physically special point, the matrix element approached zero in this limit. Note also that while $\sigma \sim C_{ww}$ had an IR divergence in the above, its $u$-derivative is essentially $N_{ww}$ and no longer does, precisely as explained in Section \ref{sec:detectors}.

The vacuum two-point function is a more non-trivial result and can be computed from the ingredients we have already assembled. It follows from Eqns. \eqref{hh_asymptotic_2pt} and \eqref{asymptotic_scalar_2pt_wightman}, and agrees with for example \cite{Liu:2023gwa},
\begin{equation}
\braket{0|\sigma(u;w) \sigma(u';w')|0} = \kappa^2 P_{ww,w'w'} (\gamma + \log (i\Lambda(u-u')))\delta^{(2)}_{S^{2}}(z,z'),
\label{celestial_graviton_2pt}
\end{equation}
where $P_{\mu\nu\rho\sigma}$ is the tensor in the graviton propagator and $\Lambda$ is an IR regulator. This is IR divergent even though this can be obtained as a limit of $\sigma$, an observable. By contrast, a finite quantity is 
\begin{equation}
\braket{0|\partial_u \sigma (u;w) \partial_u' \sigma (u';w')|0} = \kappa^2 \frac{P_{ww,w'w'}}{(u-u')^2}\delta^{(2)}_{S^2}(z,z'),
\end{equation}
which is IR-finite and corresponds to the two-point function $\braket{0|N_{ww} N_{w'w'}|0}$. This again is precisely what we found in Section \ref{sec:detectors}: this asymptotic limit of the vacuum two-point function of local fields has an IR divergence, but the derivatives of these fields have an IR finite two-point function.

\subsection{Shapiro time delay}
\label{sec:shapiro}

Our bulk definitions of arrival times in Section \ref{sec:arrival_time} smoothly match onto the familiar asymptotic notion of a Shapiro time delay. This will give a direct, perturbatively-correctable definition of the Shapiro time delay in quantum gravity and a physically transparent way of studying IR divergences in time delays as independently as possible from the massless particle, and without needing to detour through the full scattering process. That said, defining such time delay operators from asymptotic observables as discussed in Section \ref{sec:detectors} may still be fruitful.

We will identify the leading large-$L$ behavior of the arrival time observables in Section \ref{sec:arrival_time} as asymptotic time delays. In the configuration we used in Section \ref{sec:arrival_time}, this limit sends worldlines in opposite directions, symmetrically about the origin. We will use this setup as a model of extending the null rays to past and future null infinity. In this setup, we identify a Shapiro time delay operator $\tau_S$ as
\begin{equation}
\tau_{S} \equiv \lim_{L \rightarrow \infty} (\tau_2-\tau_1 - L).
\end{equation}
This captures the change in time delay from the classical value, $L$, which is initial data and so is gauge invariant. In Section \ref{sec:arrival_time}, we showed that in synchronous gauge, this is precisely the usual expression for the Shapiro time delay when the standard definitions are valid, 
\begin{equation}
\tau_S =u = \frac{\kappa}{2} \int_{-\infty}^\infty d\eta h_{zz} (x^{(0)}(\eta;u_{0}))+ O(h^2),
\end{equation}
where $h$ is operator-valued. We use light cone coordinates and the null ray is taken to propagate in the $z$-direction without loss of generality. One can compute correlators and matrix elements of this operator in the graviton Hilbert space. We will study the leading quantum gravity correction to its two-point function. We will encounter the well-documented phenomenon that the Shapiro time delay fluctuates in quantum gravity.

\begin{figure}[!htp]
\centering
\begin{tikzpicture}[scale=3.4]

% Diamond vertices
\coordinate (T) at (0,  1);
\coordinate (B) at (0, -1);
\coordinate (L) at (-1, 0);
\coordinate (R) at ( 1, 0);

% Draw diamond
\draw[thick] (T) -- (R) -- (B) -- (L) -- cycle;

% Left arc of the lens (bulges left)
\draw[thick] (B) .. controls (-0.42,-0.42) and (-0.42, 0.52) .. (T);

% Right arc of the lens (bulges right, but stays inside diamond)
\draw[thick] (B) .. controls ( 0.42,-0.52) and ( 0.42, 0.52) .. (T);

\coordinate (F) at (-0.27, -0.18-0.2);
\coordinate (G) at ( 0.30, 0.128+ 0.28-0.2);
\coordinate (E) at (0.276, -0.18-0.2);

\filldraw[black] (F) circle (0.4pt);
\filldraw[black] (G) circle (0.4pt);
\filldraw[black] (E) circle (0.4pt);

% Wiggly black line from F to G
\draw[black, line width=0.6pt, decorate, 
      decoration={snake, amplitude=1.4pt, segment length=8pt}] 
      (F) -- (G);  
      
\draw[black, thick, dashed]
    (L) .. controls (-0.5, -0.54) and (0.5, -0.54) .. (R);

% Curved double-sided arrow from F to E, same curvature as dashed line but below
% The dashed curve control points give a sag of about -0.54 at center
% We shift down by ~0.05 and restrict to the x-range of F to E
\draw[<->, black,thick]
    (-0.23, -0.44) .. controls (-0.05, -0.46) and (0.05, -0.46) .. (0.23, -0.44);

% Label 2T below the midpoint of the arrow
\node[below, font=\small] at (0, -0.45) {$L$};
\end{tikzpicture}
\caption{Penrose diagram  of the Minkowski space background without metric fluctuations. The wavy line corresponds to a light ray emitted from the $x_1$ worldline and detected by the $x_2$ worldline. The dashed curve denotes constant $t$ slices. This defines a form of arrival time captured by finite-distance measurements. The timelike worldlines are a coordinate distance $L$ apart, symmetrically about the origin. Sending $L \rightarrow \infty$ then defines a Shapiro time delay operator in quantum gravity.}
\label{fig:two_rays}
\end{figure}
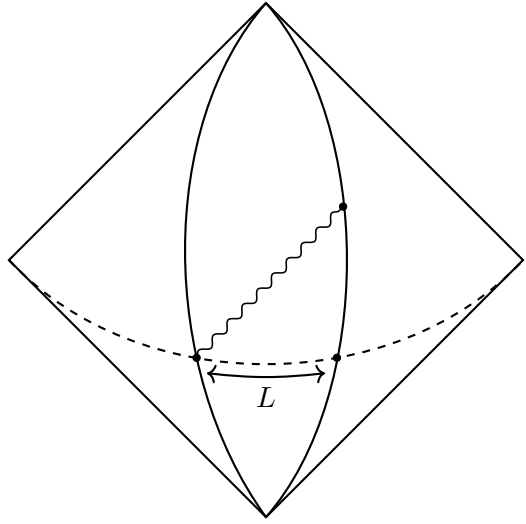

The two-point function of $u$, given in \eqref{uu_wightman_function}, is an integral over the graviton two-point function. We use the transverse-traceless two-point function (see \cite{Carney:2024wnp}), where the Wightman function is defined via $(t'-t) \rightarrow (t'-t)-i\varepsilon$. For convenience we parametrize the separation of the light ray in terms of $t$, where $t= \left(u'_{0}-u_{0}\right)$. Then the two-point function at finite $T = L/2$ is
\begin{equation}
\braket{u(u'_0) u(u_0)}_c 
= \frac{\kappa^{2}}{192 \pi^2}\int_0^\infty dk \frac{e^{-i k t - k\varepsilon}}{k^4 T^3}(4 k T(8 k^2 T^2 - 3) + 3 \sin(4 k T)).
\end{equation}
Performing the integral and taking the large $T$ limit $T\gg t>0$,
\begin{equation}
\langle u(u'_0)u(u_0)\rangle_{c} =  \frac{\kappa^{2}}{36\pi^2}(-11 + 12\log 2 + 6 \log (T/t) - 3i\pi) + O(t/T),
\label{retarded_time_shift_correlator}
\end{equation}
with commutator 
\begin{align}
\langle [u(u'_0),u(u_0)]&\rangle= -i \frac{\kappa^{2}}{6\pi} + O(t/T).
\end{align}
In the large $T$ limit, the Wightman function \eqref{retarded_time_shift_correlator} diverges logarithmically, while the commutator is finite. For earth-based detectors the logarithmic growth is negligible. The $\kappa^{2}$ prefactor makes such fluctuations far too small to be measurable \cite{Carney:2024wnp}. Strictly speaking, however, the presence of this IR divergence signifies that $u(u_0)$ is not a well-defined asymptotic observable, even though the finite-distance version is a perfectly good observable. This was inevitable because the only dimensionless quantity is $T/t$, and so the short-distance divergence in $t$ implies the function has a large-$T$ divergence. The commutator, on the other hand, has no such divergence, since by definition it vanishes at coincident points.

We have considered the large-$L$ limit of the $\tau_2-\tau_1$ observable because it recovers the standard Shapiro time delay. Even in the classical case, Shapiro time delays are known to have IR divergences, see for example \cite{Gonzo:2024xjk}. In our case, just as in the classical case, differences of time delays will end up having an IR-finite large-$L$ limit.

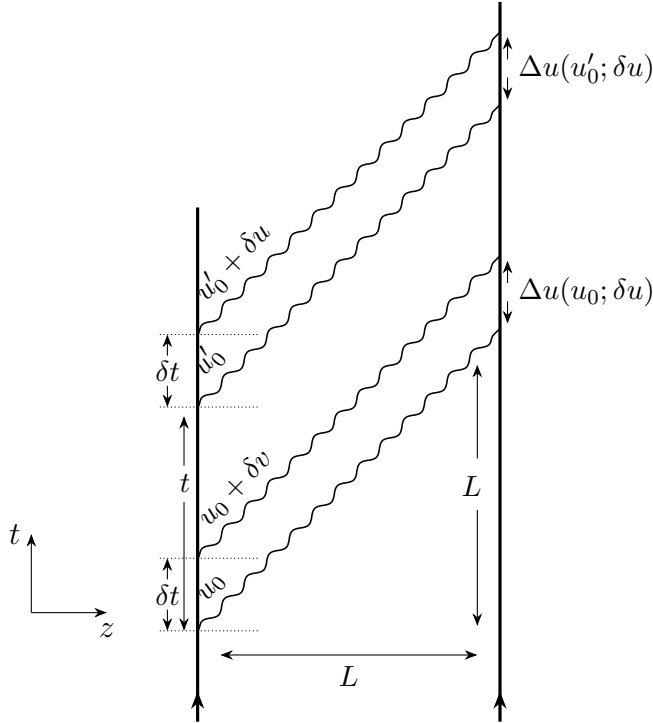
\begin{figure}[h]%[!htp]
\centering
\begin{tikzpicture}[>=Stealth, yscale=0.8]

% === Parameters ===
\def\xL{2.5}
\def\xR{6.5}
\def\ybot{2}
\def\ytop{14.5}
\def\gap{0.5}

\pgfmathsetmacro{\dx}{\xR-\xL} % = 4.5
\def\dy{5.0}                    % vertical rise

% Bottom pair emissions (at left wall)
\def\yA{1.0+2.5}       % v_0 emission
\def\yB{2.2+2.5}       % v_0 + dv emission

% Top pair emissions (at left wall)
\def\yC{7.2}       % v_0' emission
\def\yD{8.4}       % v_0' + dv emission

% Arrival points at right wall
\pgfmathsetmacro{\yAR}{\yA+\dy}
\pgfmathsetmacro{\yBR}{\yB+\dy}
\pgfmathsetmacro{\yCR}{\yC+\dy}
\pgfmathsetmacro{\yDR}{\yD+\dy}

\pgfmathsetmacro{\ang}{atan(\dy/\dx)}

% === Two thick vertical worldlines ===
\draw[very thick] (\xL, \ybot) -- (\xL, \ytop-4);
\draw[very thick] (\xR, \ybot) -- (\xR, \ytop-0.6);

% === Upward arrows at bottom of vertical lines ===
\draw[->, thick] (\xL, \ybot) -- (\xL, \ybot+0.5);
\draw[->, thick] (\xR, \ybot) -- (\xR, \ybot+0.5);

% === Diagonal light rays (wiggly) ===
\draw[line width=0.6pt, decorate, decoration={snake, amplitude=1.4pt, segment length=12pt}] (\xL, \yA) -- (\xR, \yAR);
\draw[line width=0.6pt, decorate, decoration={snake, amplitude=1.4pt, segment length=12pt}] (\xL, \yB) -- (\xR, \yBR);
\draw[line width=0.6pt, decorate, decoration={snake, amplitude=1.4pt, segment length=12pt}] (\xL, \yC) -- (\xR, \yCR);
\draw[line width=0.6pt, decorate, decoration={snake, amplitude=1.4pt, segment length=12pt}] (\xL, \yD) -- (\xR, \yDR);

% === Horizontal dotted lines ===
\draw[densely dotted, thin] (\xL-0.5, \yA) -- (\xL+0.8, \yA);
\draw[densely dotted, thin] (\xL-0.5, \yB) -- (\xL+0.8, \yB);
\draw[densely dotted, thin] (\xL-0.5, \yC) -- (\xL+0.8, \yC);
\draw[densely dotted, thin] (\xL-0.5, \yD) -- (\xL+0.8, \yD);

% === Ray labels ===
\path (\xL, \yA) -- (\xR, \yAR)
  node[pos=0.1, above=0pt, rotate=\ang, anchor=south, font=\small] {$u_0$};
\path (\xL, \yB) -- (\xR, \yBR)
  node[pos=0.18, above=0pt, rotate=\ang -5, anchor=south, font=\small] {$u_0+\delta v$};
\path (\xL, \yC) -- (\xR, \yCR)
  node[pos=0.1, above=0pt, rotate=\ang, anchor=south, font=\small] {$u_0'$};
\path (\xL, \yD) -- (\xR, \yDR)
  node[pos=0.18, above=0pt, rotate=\ang-5, anchor=south, font=\small] {$u'_0+\delta u$};

% === delta t bottom (between v_0 and v_0+dv emissions at left wall) ===
\draw[->, thin] ({\xL-0.4}, {\yA+0.35}) -- ({\xL-0.4}, {\yA});
\draw[->, thin] ({\xL-0.4}, {\yA+0.4+0.45}) -- ({\xL-0.4}, {\yB});
\node[right, font=\small] at ({\xL-0.7},\yA+0.6) {$\delta t$};

% === delta t top (between v_0' and v_0'+dv emissions at left wall) ===
\draw[->, thin] ({\xL-0.4}, {\yC+0.35}) -- ({\xL-0.4}, {\yC});
\draw[->, thin] ({\xL-0.4}, {\yC+0.4+0.45}) -- ({\xL-0.4}, {\yD});
\node[right, font=\small] at ({\xL-0.7},\yC+0.6) {$\delta t$};

% === t between groups (left side, from yB to yC) ===
\pgfmathsetmacro{\tmid}{(\yB+\yC)/2}
\draw[<-, thin] ({\xL-0.18}, {\yA}) -- ({\xL-0.18}, {\tmid-0.32});
\draw[->, thin] ({\xL-0.18}, {\tmid-0.32+0.6}) -- ({\xL-0.18}, {\yC-0.15});
\node[right, font=\small] at ({\xL-0.38}, \tmid) {$t$};

% === 2T vertical on right side (from yA dotted line up to yAR) ===
\pgfmathsetmacro{\twoTmid}{(\yA+\yAR)/2}
\draw[<-, thin] ({\xR-0.3}, {\yA+0.08}) -- ({\xR-0.3}, {\twoTmid-0.5});
\draw[->, thin] ({\xR-0.3}, {\twoTmid+0.35}) -- ({\xR-0.3}, {\yAR-0.6});
\node[right, font=\small] at ({\xR-0.65}, \twoTmid-0.1) {$L$};

% === 2T horizontal at bottom ===
\draw[<->, thin] (\xL+0.3, \yA-0.4) -- (\xR-0.3, \yA-0.4);
\node[below, font=\small] at ({(\xL+\xR)/2},\yA -0.4) {$L$};

% === Delta v(v_0; delta v) on right side ===
\pgfmathsetmacro{\dvmid}{(\yAR+\yBR)/2}
\draw[<-, thin] ({\xR+0.1}, {\yAR+0.08}) -- ({\xR+0.1}, {\dvmid-0.15});
\draw[->, thin] ({\xR+0.1}, {\dvmid+0.15}) -- ({\xR+0.1}, {\yBR-0.08});
\node[right, font=\small] at ({\xR+0.1}, \dvmid) {$\Delta u(u_0;\delta u)$};

% === Delta v(v_0'; delta v) on right side (top pair) ===
\pgfmathsetmacro{\dvpmid}{(\yCR+\yDR)/2}
\draw[<-, thin] ({\xR+0.1}, {\yCR+0.08}) -- ({\xR+0.1}, {\dvpmid-0.15});
\draw[->, thin] ({\xR+0.1}, {\dvpmid+0.15}) -- ({\xR+0.1}, {\yDR-0.08});
\node[right, font=\small] at ({\xR+0.1}, \dvpmid) {$\Delta u(u_0';\delta u)$};

% === Axes (t, z) in bottom-left ===
\coordinate (O) at (0.3, 3.8);
\draw[->] (O) -- ++(0, 1.3) node[left] {$t$};
\draw[->] (O) -- ++(1, 0) node[below] {$z$};
\end{tikzpicture}
\caption{Here, we show the setup for computing the two-point function, Eq. \eqref{finite_shift_correlator}, of the finite-distance observable $\Delta u$ that has a finite large-$L$ limit. The wavy lines denote null geodesics. We define $t=(u'_0-u_0)$ and $\delta t=\delta u$. Here we depict the configuration $0<\delta t<t$. The computation is carried out in synchronous gauge.}
    \label{fig:light-rays}
\end{figure}
As shown in Appendix \ref{app:arrival_time_arbitrary_gauge}, the difference in the arrival times of two light rays with different emission times can be written as
\begin{equation}
\Delta u (u_{0}; \delta u) \equiv u(u_{0}+\delta u)- u(u_{0}).
\label{finite_shift_obv}
\end{equation}
See Fig. \ref{fig:light-rays} for this configuration. This quantity will be finite in the large-$T$ limit. As explained in \cite{Camanho:2014apa}, a series of shocks can accumulate time delays and could perhaps accumulate coherently to lead to an infinite time delay. Here we have vacuum fluctuations instead of shocks, and this type of accumulation does not occur in our computation of this observable.

The connected correlator for the observable $\Delta u$ is  \begin{equation}
\langle \Delta u (u'_{0}; \delta u)\Delta u (u_{0}; \delta u) \rangle_{c}
= \frac{\kappa^{2}}{48 \pi^2}\int_{0}^{\infty}dk\, 
\frac{e^{-ikt - k\varepsilon}}{k^{4}T^{3}} \sin^{2}\!\left(\frac{k \delta u}{2}\right)
\left[4kT(8 k^2 T^2 -3)+ 3\sin(4k T)\right],
\label{finite_shift_correlator}
\end{equation}
where $t=(u'_0 -u_0)$. Furthermore, in the large-$T$ limit in question, $T\gg t>0$, the correlator \eqref{finite_shift_correlator} is finite,
\begin{equation}
\langle \Delta u (u'_{0}; \delta u)\Delta u (u_{0}; \delta u) \rangle_{c} = \frac{\kappa^{2}}{6\pi^2}\log \left(1- \frac{ \delta t^2}{t^{2}}\right)+ O(t/T),
\label{finite_shift_correlator_case_1}
\end{equation}
where $\delta t=\delta u$, and we work in the regime $0<\delta t<t$. 

Another interesting case is when $0<t<\delta t$. In the same large $T$ limit, $T\gg \delta t$, the correlator \eqref{finite_shift_correlator} is
\be
\langle \Delta u (u'_{0}; \delta u)\Delta u (u_{0}; \delta u) \rangle_{c} = \frac{\kappa^{2}}{6\pi^2}\left[-i\pi+ \log \left(\frac{ \delta t^2}{t^{2}}-1\right)\right]+ O(\delta t/T).
\ee
This quantity remains finite at large $T$, but grows  logarithmically with $\delta t$. In this sense, vacuum fluctuations of the metric can accumulate Shapiro time delays logarithmically in time, somewhat reminiscently of memory.

A fruitful next step would be to relate the two point function of time delays to an underlying asymptotic four point function of operator insertions at $\scri^+$ and $\scri^-$, or to the S-matrix with gravitational interactions. We take initial steps towards this in Section \ref{sec:detectors}, but leave this goal to future work.

Finally, we conclude by exploring the graviton matrix element of the Shapiro time delay. We again define this via the large-$L$ limit of the finite-distance matrix element, with a regulator $\varepsilon$ to define this limit,
\begin{align}
\braket{0|\tau_S|h(k)} &= \frac{\kappa }{2} P_{z\rho}P_{z\sigma} \epsilon^{\rho \sigma} \int_{-\infty}^{\infty} d\eta e^{ik^\mu x^{(0)}_\mu(\eta;u_0) - |\eta| \varepsilon }
\nonumber
\\
&=
\kappa \pi  P_{z\rho}P_{z\sigma} \epsilon^{\rho \sigma} e^{-i k^0 u_0} \delta ( k^0-k^z).
\end{align}
Here we use $x^{(0)}(\eta;u_0) = (\eta+u_0,0,0,\eta)$. The delta function enforces $k^\mu = (k^0,0,0,k^0)$ which gives a zero matrix element. 

However we can consider a half-infinite Shapiro time delay quantity $\tau_{S}'$ instead, which models taking only one worldline to infinity,
\begin{align}
\braket{0|\tau_S'|h(k)} &= \frac{\kappa  }{2} P_{z\rho}P_{z\sigma} \epsilon^{\rho \sigma} \int_{0}^{\infty} d\eta e^{ik^\mu x^{(0)}_\mu(\eta;u_0) - \eta \varepsilon}
\nonumber
\\
&=
-\frac{i\kappa }{2} P_{z\rho}P_{z\sigma} \epsilon^{\rho \sigma} e^{-i k^0 u_0} \frac{1}{k^0-k^z-i\varepsilon}.
\end{align}
This is now non-zero. It lacks any collinear singularity.

\subsection{Round trip time in the IR}

We discuss infrared behavior of the elapsed proper time interferometer observable in \cite{Carney:2024wnp}. Taking $L$ large makes the interferometer arbitrarily large. We will address two salient features of this configuration, the large-$L$ divergence of its two-point function, and the interpretation of $\tau$ as an operator on the celestial sphere.

We can analyse the large-$L$ limit of $\tau$ via the two point function,
\begin{equation}
\braket{\tau(t) \tau(0)}_c = 
\frac{\kappa^2}{2\pi^2 }\int_{0}^{\infty}dk e^{-i k t- k\varepsilon} \bigg[\frac{3 - \cos(2 k L)}{6k}  - \frac{3 + \cos(2 k L)}{2 k^3 L^2} + \frac{\sin(2k L)}{k^4 L^3}\bigg].
\end{equation}
We can perform the integral and take the large $L$ limit $L\gg t>0$. We emphasize that this is different from the usual asymptotic limit in which interferometers are far away from some classical source. Ostensibly, this limit amounts to the interferometer asymptotically occupying a slice of the entire Penrose diagram of flat space, including timelike, spatial, and null infinity, though we will return to this point shortly. For now, we will let direct computation guide us. The two-point function becomes
\begin{equation}
\langle \tau(t)\tau(0) \rangle_{c}  = \frac{\kappa^2}{72\pi^2 }\left(-34+18\log(2)+18 \log(L/t) - 9i \pi \right) + O(t/L).
\label{time_shift_correlator_asymptotic}
\end{equation}
We obtain a form similar to Eq. \eqref{retarded_time_shift_correlator} under the replacement $T = L/2$. Note that direct computation shows that at large $t$, $\braket{\tau(t) \tau(0)}_c \sim 
 -\kappa^2(2L^2)/(15 \pi^2 t^2)$, essentially because in this regime, the two-point function of $h$ determines the large-$t$ decay.

The divergence in Eq. \eqref{time_shift_correlator_asymptotic} is worth discussing. In the regime $L\gg t$, the correlator $\braket{\tau(t)\tau(0)}_c$ diverges in the large-$L$ limit and also equivalently in the small-$t$ limit via the single dimensionless combination $t/L$. However, as in \cite{Carney:2024wnp},  we can smear $\tau$ over a finite-time window of width $T_w$, which models a finite-time observation, and this removes the small-$t$ divergence. Doing so, we find the large-$L$ divergence nevertheless remains, now of the form $\log(L/T_w)$. From this, we conclude that even when we account for finite-time resolution, it seems unclear whether $\tau$ is formally a good asymptotic observable when asymptotic infinity is reached by sending $L \rightarrow \infty$. 

In synchronous gauge, the metric fluctuations lie entirely along null infinity, rather than in the bulk where gravity is genuinely dynamical. However there is of course no requirement that every finite distance observable must have a well-behaved asymptotic limit, so the large-$L$ divergence here does not imply an obvious contradiction or invalid ingredient in the setup. In fact, in the case of the Shapiro time delays, we found that $\tau_2-\tau_1$ was indeed IR divergent, while differences of this quantity were IR finite. The analogous quantity, $\braket{(\tau(t)-\tau(t+\delta t)) (\tau(0)-\tau(\delta t))}$, is also finite in the large-$L$ limit. From this point of view, the large-$L$ divergence in the round trip time observable appears consistent with EFT expectations, appearing to confirm the conclusion of \cite{Carney:2024wnp}.

Despite the large-$L$ divergence, we can anticipate the operator on the celestial sphere that $\tau$ approaches in the large-$L$ limit. As we have explained, the round trip time $\tau$ can be written as
\begin{equation}
\tau=\tau_3-\tau_2 + \tau_2-\tau_1.
\end{equation}
This suggests a strategy for identifying its asymptotic limit, which we sketch. Both $\tau_3-\tau_2 - L $ and $\tau_2-\tau_1 - L$ individually are Shapiro time delay operators, as explained in Section \ref{sec:shapiro}. In the large-$L$ limit, these two Shapiro time delay operators are pushed to asymptotic infinity. In synchronous gauge, one expects they become Shapiro time delay operators of the form $\tau_S^{(1)} \sim  \int d\eta \, h_{nn}(\eta) \sim \int du \, C_{ww}(u)$ integrated over $\scri^-$ or $\scri^+$ for the outgoing or incoming ray respectively, where here $n$ is taken tangent to the celestial sphere, and is represented on the celestial sphere by $w$. One possible subtlety is that it is crucial the null ray reflects off a timelike geodesic and changes directions, which at asymptotic infinity may involve incorporating spacelike infinity. We leave further analysis to future work, including the implications for  connections between interferometer measurements and the celestial sphere, as explored recently in \cite{He:2025hag, Ciambelli:2025fbo, Ciambelli:2026pwi} and related work.

\section{Discussion}

Even when spacetime fluctuates, forms of geometry persist. When quantum gravity effects are weak, it is hardly surprising that classical measures of geometry acquire the hallmark correlations of quantum operators. In this paper, we showed how this can happen using insights from the study of gravitationally dressed observables. We built geometric observables that model measurements and experimental setups. We further showed how these same observables matched onto memory and Shapiro time delay observables on the celestial sphere. Our analysis focused on leading graviton corrections, which we computed in closed form. Our results link a number of topics, including celestial holography, gravitational dressing, and experimental probes of the interface between gravity and quantum mechanics. We expect that higher orders and non-vacuum  states can be included via the better-studied calculus of dressed local operators, for instance see \cite{Giddings:2025xym, Cheung:2026euf}. 

One broad aim of this work is to show how to study geometric observables by using common tools in perturbative quantum gravity. Further computing graviton corrections to measures of geometry may probe a number of interesting ideas, which we comment on briefly.

Studying simple quantities like distance and time allowed us to make contact with basic properties of geometric observables, and illustrate how to compute them. We expect our approach can also be used to include graviton effects in more complex geometric measures like horizon areas, extremal surface areas, and  focusing. These quantum-corrected geometric measures arise in extensions of quantum extremal surfaces \cite{Colin-Ellerin:2025dgq} and quantum focusing \cite{Bousso:2015mna} into the regime of perturbative quantum gravity. Studying graviton corrections to geometric observables may shed light on the detailed properties of subregions beyond AdS/CFT, including features like their entanglement wedges \cite{Bousso:2022hlz}.

General principles of quantum gravity may be challenging to formulate precisely, but in classical gravity, they are mainly stated in terms of geometry. So developing geometric observables may lead to novel objects that sharply capture properties of quantum gravity. An obvious target is causality, for example as in Gao-Wald \cite{Gao:2000ga}, in the presence of gravitons. Causality also places bounds on higher-derivative couplings, for instance as in \cite{Camanho:2014apa, Afkhami-Jeddi:2016ntf,  Hartman:2022njz}.

Studying finite-distance geometric observables may further illuminate operations on the celestial sphere. While the large-radius limit of bulk operators is well-studied, we expect the finite-distance observables naturally generate the $1/R$ corrections, which may lead to additional operators on the celestial sphere and enter into subleading soft behavior \cite{Himwich:2019dug, Cachazo:2014fwa}. Memory is a classical phenomenon, and so further connecting experimental measurements of distance to operators on the celestial sphere may clarify the operational meaning of quantum extensions of memory, as in \cite{Moult:2025njc}. More broadly, identifying asymptotic observables \cite{Caron-Huot:2023vxl, Caron-Huot:2025ymc} that compute the result of finite-distance experiments in the presence of gravitons may bring us closer to defining geometric observables solely from spacetime boundaries, and ultimately to exploring non-perturbative regimes.

\begin{center} 
{\bf Acknowledgments}
\end{center}

We thank 
Leonardo Badurina, Yanbei Chen, Daniel Carney, Clifford Cheung, Daine Danielson, John Donoghue, Steven Giddings, Temple He, Philipp Hoehn, Joon-Hwi Kim, Per Kraus, Aaron Markowitz, Prahar Mitra, Ian Moult, Brett Oertel, Rodolfo Russo, Gautam Satishchandran, Brian Seymour, Antony Speranza, Sander Vermeulen, Jeffrey Wack, Colin Weller, Jordan Wilson-Gerow, Lihang Zhou, and Kathryn Zurek for discussions and/or comments on the draft. We thank Daniel Carney for sharing his unpublished notes on length and time correlators and connections to interferometry, as well as initial collaboration on these topics. During this work, A. Sivaramakrishnan was supported by the Heising-Simons Foundation “Observational Signatures of Quantum Gravity” Collaboration Grant No. 2021-2817, the DOE Early Career Award DE-SC0025581, the Sloan Foundation, the Templeton Foundation grant 63670 on “Emergence of time from chaos in operator algebras”, and DOE Grant DE-SC0025547. M. Kolo\u{g}lu was supported by the Yale Mossman Prize Postdoctoral Fellowship as well as by the UK EPSRC grant EP/W019663/1, ``CFT and Gravity: Heavy States and Black Holes''. M. Karydas was supported by the U.S. Department of Energy, Office of High Energy Physics, under Contract No. DE-AC02-05CH11231 and by the DOE Quantum Information Science Enabled Discovery (QuantISED) for High Energy Physics grant KA2401032, as well as by the Heising-Simons Foundation under grant 2023-4467, “Testing the Quantum Coherence of Gravity''.

\newpage 

\appendix

\section{Geometric observables in an arbitrary gauge}

\label{app:geometric_observables_arbitrary}

Geometric observables are storied objects in general relativity. Despite this, they are often studied on a case-by-case basis, with less attention paid to common structure. Computations make heavy use of certain gauges and are phrased in terms of details specific to the relevant experimental setups. Diffeomorphism invariance is widely understood, although can be far from manifest in the actual expressions presented. Geometric observables are often discussed separately from curvature observables, although they are understood to encode the same information. In this Appendix, we present a unified treatment of length and time observables that nicely sets up our extensions to perturbative quantum gravity. In particular, detailed control over gauge invariance at the level of states and operators, for instance in the form of Ward identities, plays an enormous role in gauge theory computations. So, one goal of our presentation will be to make diffeomorphism invariance manifest and easily verified in both the classical and quantum cases. We will also discuss the initialization of the worldlines used to dress these observables. The material in this section is partly new and partly a synthesis of well-known ideas in the literature, including some that are broadly known even if not explicitly stated in the form we present. In the main body, we make use of the results of this Appendix for extensions to quantum gravity.

\subsection{General derivation}

\label{app:geometric_observables_arbitrary_derivation}
 
In this section we show that the geometric observables we study take a particularly simple form in an arbitrary gauge. Remaining gauge-agnostic exposes certain relations between the observables in the purely classical setting. Our strategy will be to first compute $(d/d\tau)^2 O_g$ and then integrate to find $O_g$. This derivation is far more concise, transparent, and amenable to higher-order corrections than directly deriving the $O_g$ expressions we find.

As in the body of this paper, we consider the timelike geodesics $x_1, x_2$ with the following initial data, imposed at some initial time $\tau_\Sigma$, which we will take to be $-\infty$:
\begin{align}
& x_2^{(0),\mu} (\tau_{\Sigma}) - x_1^{(0),\mu}(\tau_{\Sigma}) = L n^\mu \equiv L (0,\mathbf{n}),
\\
& \dot{x}_1^{(0),\mu} (\tau_{\Sigma}) = \dot{x}_2^{(0),\mu}(\tau_\Sigma) = \delta^{\mu}_{0}.
\end{align}

In this setup, we will compute the proper length $\Omega(\tau_1,\tau_2)$ of the connecting geodesic $x(\alpha)$ between the points $x_1(\tau_1)$ and $x_2(\tau_2)$. The geodesic $x$ can be a spacelike, timelike, or null geodesic. The boundary conditions of the connecting geodesic $x$ are $x(0) =x_1(\tau_1)$ and $x(1)=x_2(\tau_2)$. The proper length of $x$ is 
\begin{align}
\Omega(\tau_1,\tau_2) &=\int d\alpha \sqrt{(\eta_{\mu\nu}+\kappa h_{\mu\nu})(\dot{x}^{(0),\mu}+\dot{x}^{(1),\mu}) (\dot{x}^{(0),\nu}+\dot{x}^{(1),\nu})}
\nonumber
\\
&=
L_\tau + \frac{1}{L_\tau }\int d\alpha \, \dot{x}^{(1)} \dot{x}^{(0)} 
+\frac{\kappa}{2L_\tau}\int d\alpha \, h_{\mu\nu}\dot{x}^{(0),\mu}\dot{x}^{(0),\nu},
\end{align}
where $L_\tau = \sqrt{L^2-(\tau_2-\tau_1)^2}$ and 
\begin{equation}
x^{(0),\mu} = (\tau_1 + \alpha(\tau_2-\tau_1), \alpha L \mathbf{n}), ~~~~~ \dot{x}^{(0),\mu} = (\tau_2-\tau_1, L \mathbf{n}),
\end{equation}
with $n^\mu = (0,\mathbf{n})$. Expanding
$\Omega = \Omega^{(0)}+\Omega^{(1)}+\ldots$ we have
\begin{equation}
\Omega^{(1)} = \frac{1}{L_\tau } \dot{x}^{(0)}(x^{(1)}(1)-x^{(1)}(0))
+
\frac{\kappa}{2L_\tau}\int d\alpha \, h_{\mu\nu}\dot{x}^{(0),\mu}\dot{x}^{(0),\nu},
\end{equation}
which more explicitly is
\begin{align}
\Omega^{(1)} =& \frac{1}{L_\tau } \left(L (x_2^{(1),n}(\tau_2)-x^{(1),n}_1(\tau_1)) -(\tau_2-\tau_1) (x^{(1),0}_2(\tau_2)-x^{(1),0}_1(\tau_1))
\right)
\nonumber
\\
&+
\frac{\kappa}{2L_\tau}\int d\alpha \, ( (\tau_2-\tau_1)^2 h_{00}+2L(\tau_2-\tau_1) h_{0n}+L^2 h_{nn} )\,.
\end{align}
We will now take derivatives to compute $\ddot{\Omega}$. We use the derivative $\partial_\tau \equiv \partial_{\tau_1} + \partial_{\tau_2}$ to compute how $\Omega$ changes as we vary both $\tau_1$ and $\tau_2$ simultaneously. We will need the following derivatives,
\begin{equation}
\partial_\tau x^{(0),\mu} = (1,\mathbf{0}),~~~\partial_\tau \dot{x}^{(0),\mu} =0.
\end{equation}
We will also need the geodesic equation,
\begin{equation}
\frac{d^2}{d\tau_i^2} x^{(1),\mu}_i =- \frac{\kappa}{2}\eta^{\beta \mu} (2\partial_0 h_{0\beta}-\partial_\beta h_{00}).
\end{equation}
Now we determine the second derivative of $\Omega$,
\begin{align}
\frac{d^2}{d\tau^2}\Omega^{(1)} =&  \frac{\kappa}{L_\tau } \left(-\frac{1}{2}L (2\partial_0h_{0n}-\partial_n h_{00})
-(\tau_2-\tau_1)\frac{1}{2} \partial_0h_{00}
\right)\bigg|^{x_2^{(0)}(\tau_2)}_{x_1^{(0)}(\tau_1)}
\nonumber
\\
&+
\frac{\kappa}{2L_\tau}\int d\alpha \, \left( (\tau_2-\tau_1)^2 \partial_0^2 h_{00}+2L(\tau_2-\tau_1) \partial_0^2 h_{0n}+L^2 \partial_0^2 h_{nn} \right),
\end{align}
where in the last line, we used that $\partial_\tau x^{(0),\mu} = (1,\mathbf{0})$ implies that $\partial_\tau h_{\mu\nu}(x^{(0),\mu}) = \partial_0 h_{\mu\nu}(x^{(0),\mu})$ by the chain rule. Similarly,
\begin{equation}
\frac{d}{d\alpha}h_{\mu\nu}(x^{(0)}(\alpha)) =((\tau_2-\tau_1) \partial_0 + L \partial_n)h_{\mu\nu}(x^{(0)}(\alpha))
\end{equation}
allows us to recombine the terms as
\begin{align}
\frac{d^2}{d\tau^2}\Omega^{(1)} = \frac{\kappa}{2L_\tau } \int d\alpha \,& \bigg(
((\tau_2-\tau_1) \partial_0 + L \partial_n)\left(
-2L\partial_0h_{0n}+L\partial_n h_{00}
-(\tau_2-\tau_1) \partial_0h_{00}\right)
\\
&~~
+(\tau_2-\tau_1)^2 \partial_0^2 h_{00}+2L(\tau_2-\tau_1) \partial_0^2 h_{0n}+L^2 \partial_0^2 h_{nn} \bigg).
\end{align}
This simplifies to
\begin{equation}
\frac{d^2}{d\tau^2}\Omega^{(1)} = \frac{\kappa}{2L_\tau } \int d\alpha \, \left(
-2L^2\partial_n \partial_0 h_{0n}+L^2 \partial_n^2 h_{00}
+L^2 \partial_0^2 h_{nn} \right).
\end{equation}
In terms of the linearized Riemann tensor,
\begin{equation}
R^{(1)}_{0n0n} =\kappa( \partial_0 \partial_n h_{0n} - \frac{1}{2}\partial_n^2 h_{00} - \frac{1}{2} \partial_0^2 h_{nn}),
\end{equation}
we have
\begin{equation}
\frac{d^2}{d\tau^2}\Omega^{(1)} = -\frac{L^2}{L_\tau } \int_0^1 d\alpha \, R^{(1)}_{0n0n}(x^{(0)}(\alpha)).
\label{OmegaRiemannSecondDerivativeForm}
\end{equation}
To solve this one must specify boundary conditions, which we do via the $i\varepsilon$ prescription $d/d\tau \rightarrow d/d\tau +\varepsilon$. This gives
\begin{equation}
\Omega^{(1)} = -\frac{L^2}{L_\tau } \int_{-\infty}^\tau d\tau' (\tau-\tau') \int_0^1 d\alpha \, R^{(1)}_{0n0n}(x^{(0)}(\tau',\alpha)),
\label{OmegaRiemannForm}
\end{equation}
where $x^{(0)}(\tau',\alpha)$ is the unperturbed geodesic that connects $x_1^{(0)}(\tau')$ and $x_2^{(0)}(\tau_2-\tau_1+\tau')$. The curve $x^{(0)}(\tau',\alpha)$ is the original $x^{(0)}(\alpha)$ extended backwards in time to form a family of geodesics that connects $x_1, x_2$, parametrized by the constant proper time shift $\tau'$. We have also reproduced this formula by directly computing $\Omega$ in various cases, but do not give the rather lengthy derivation here. Shortly, we will extract distance and time observables relatively easily from Eqs. \eqref{OmegaRiemannSecondDerivativeForm} and \eqref{OmegaRiemannForm}.

\subsection{Proper distance}
\label{app:proper_distance_arbitrary_gauge}

We can determine the spacelike distance $\sigma$ from Eq. \eqref{OmegaRiemannForm}. For simplicity, set $\tau_1 = \tau_2$. Dropping the subscripts on $\tau_i$, we have (see Eq. \eqref{OmegaRiemannSecondDerivativeForm})
\begin{equation}
\ddot{\sigma}^{(1)}(\tau) = - L \int_0^1 d\alpha \, R^{(1)}_{0n0n}(x^{(0)}(\tau, \alpha)),
\end{equation}
and
\begin{equation}
\sigma^{(1)}(\tau) = - L \int_{-\infty}^\tau d\tau' (\tau-\tau') \int_0^1 d\alpha \, R^{(1)}_{0n0n}(x^{(0)}(\tau', \alpha)).
\end{equation}
We have also derived this expression directly, rather than obtaining it by integrating $\ddot{\sigma}$.

\subsection{Arrival time}
\label{app:arrival_time_arbitrary_gauge}  

There are various arrival time observables. We will analyse two in an arbitrary gauge. This allows us to identify the observable that captures a finite-distance version of the Shapiro time delay and then smoothly matches onto the expected infinite-distance quantity.

We first consider a seemingly strange observable, which is the difference between the arrival time $\tau_2$ at $x_2$ and the departure time $\tau_1$ from $x_1$. This observable does not appear to be the typical arrival time observable, as it involves $\tau_1$ and $\tau_2$, and so cannot obviously be measured along a single worldline. However, we will see the importance of this observable shortly. 

The quantity $\tau_2-\tau_1$ is determined by the condition $\Omega(\tau_1,\tau_2) = 0$, although it will be more convenient to use the squared proper length,
\begin{equation}
\Omega^2=(\Omega^{(0)}(\tau_1,\tau_2))^2 + 2\Omega^{(0)}(\tau_1,\tau_2) \Omega^{(1)}(\tau_1,\tau_2)=0,
\label{OmegaNullConstraint}
\end{equation}
where $\tau_2 = \tau_2(\tau_1)$ is a function of $\tau_1$, which we take to be input. We can expand 
\begin{equation}
\tau_2 = \tau_2^{(0)}+\tau_2^{(1)},
\end{equation}
where $\tau_2^{(0)} = \tau_1+L$. Expanding Eq. \eqref{OmegaNullConstraint} then gives
\begin{equation}
L^2-(\tau_2^{(0)}+\tau_2^{(1)}-\tau_1)^2 + 2\Omega^{(0)}(\tau_1,\tau_2^{(0)})\Omega^{(1)}(\tau_1,\tau_2^{(0)}) =0 ,
\end{equation}
or to leading order in $h$,
\begin{equation}
\tau_2^{(1)}  = \frac{1}{L} \Omega^{(0)}(\tau_1,\tau_2^{(0)})\Omega^{(1)}(\tau_1,\tau_2^{(0)}).
\end{equation}
Now we can act with $d^2/d\tau_1^2$. This is the only independent $\tau$ parameter here, but certain properties are more manifest in terms of $d/d\tau$, 
\begin{equation}
\frac{d}{d\tau} \Omega^{(0)}(\tau_1,\tau_2) = 0,~~~~ \frac{d}{d\tau} f((\tau_1-\tau_2)^2)\bigg|_{\tau_2 = \tau_2^{(0)}}=\frac{d}{d\tau_1} f((\tau_1-(\tau_1+L))^2).
\end{equation}
So we have
\begin{align}
\frac{d^2}{d\tau_1^2} \tau_2^{(1)}  &= \frac{1}{L} \frac{d^2}{d\tau^2} \left( \Omega^{(0)}(\tau_1,\tau_2)\Omega^{(1)}(\tau_1,\tau_2) \right)\bigg|_{\tau_2 = \tau_2^{(0)}}
\\
&=
\frac{1}{L}  \Omega^{(0)}(\tau_1,\tau_2) \frac{d^2}{d\tau^2} \Omega^{(1)}(\tau_1,\tau_2) \bigg|_{\tau_2 = \tau_2^{(0)}}
.
\end{align}
Noting that the prefactors of $\Omega^{(0)}(\tau_1,\tau_2^{(0)}) = L_\tau$ cancel upon the replacement, using \eqref{OmegaRiemannSecondDerivativeForm} gives, through leading order in $h$,
\begin{equation}
\frac{d^2}{d\tau_1^2} (\tau_2-\tau_1)  = 
-L \int_0^1 d\alpha \, R^{(1)}_{0n0n}(x^{(0)}(\tau_1,\alpha)),
\label{ArrivalTimeRiemann}
\end{equation}
where $x^{(0)}(\alpha)$ is the zeroth-order null ray between $x_1(\tau_1)$ and $x_2(\tau_2^{(0)})$. The solution is
\begin{equation}
\tau_2-\tau_1  =L 
-L \int_{-\infty}^{\tau_1} d\tau' (\tau_1-\tau') \int_0^1 d\alpha \, R^{(1)}_{0n0n}(x^{(0)}(\tau',\alpha)).
\label{ArrivalTimeRiemann}
\end{equation}
We have also derived this by directly computing $\tau_2$ rather than its second derivative, though we do not include the lengthy derivation here. 

We can also consider an observer measuring the difference in arrival times of two null signals along their worldline. Suppose there are two null rays arriving at the second worldline at times $\tau_2, \tau_2'$ after being emitted from the first worldline at times $\tau_1, \tau_1'$. The time difference $\tau_{2}'-\tau_{2}$ can be measured by a single observer and is finite. We can write this difference as follows,
\begin{equation}
\tau_{2}'-\tau_2 =(\tau_{2}'-\tau_{1}')-(\tau_2-\tau_1)+\tau_{1}'-\tau_1. 
\end{equation}
Once again, we see the appearance of the observable we studied earlier but now for both null rays, $\tau_2-\tau_1$ and $\tau_2'-\tau_1'$.  In this setup, $\tau_1, \tau_1'$ are inputs, and so have no metric correction. They are the times along worldline one at which the null rays are fired. This means that $\tau_2'-\tau_2$ is determined entirely by the observables we identified earlier, $\tau_2-\tau_1$ and $\tau_2'-\tau_1'$, but unlike those individual quantities, appears more obviously measurable by a single observer.

\subsection{Round trip time}
\label{app:round-trip_time_arbitrary_gauge}

Finally, we turn to the round trip time of a null ray emitted from $x_1(\tau_1)$, reflecting at $x_2(\tau_2)$ and returning to $x_1(\tau_3)$. The round trip time observable $\tau_3-\tau_1$ has been derived in various gauges, in gauge-independent form, and by explicitly constructing the intermediate null geodesic. The method of derivation we present here conveniently bypasses some of these intermediate stages and condenses previous derivations. The main observation is strikingly intuitive. It is an algebraic fact that
\begin{equation}
\tau_3-\tau_1 = \tau_+ + \tau_-,
\end{equation}
where
\begin{equation}
\tau_+ = \tau_2-\tau_1, ~~~~~~ \tau_- = \tau_3-\tau_2,
\end{equation}
each of which is given by Eq. \eqref{ArrivalTimeRiemann},
\begin{equation}
\tau_3-\tau_1 = 2L - L \int_{-\infty}^{\tau_1} d\tau'(\tau_1-\tau') \int d\alpha \, R^{(1)}_{0n0n}(x^{(0)}(\tau',\alpha)),
\end{equation}
where $x^{(0)}(\alpha)$ is the unperturbed path of the null ray,
\begin{equation}
x^{(0),\mu}(t_1,\alpha) = \theta(1-\alpha)(\alpha L +t_1,\alpha \mathbf{n} L )+ \theta(\alpha-1)(\alpha L+t_1,(2-\alpha) \mathbf{n} L),
\end{equation}
with $\alpha \in [0,2]$ and proper time $\tau_1$ expressed as coordinate time $t_1$.

We emphasize the following: $\tau_3-\tau_1$ appears to be defined along a single worldline, but it decomposes into the sum of two simpler, but more manifestly relational observables $\tau_2-\tau_1$ and $\tau_3-\tau_2$. We have also derived $\tau_3-\tau_1$ directly and confirmed that the full expression indeed decomposes into those of $\tau_{\pm}$.

\subsection{Diffeomorphism invariance}
\label{app:geometric_observables_arbitrary_diffs_and_initializations}

The formulas we have derived for various geometric observables $O_g$ make diffeomorphism invariance manifest. The diffeomorphisms we consider, and require invariance under, are those that do not act at infinity, or small diffeomorphisms. In this section, we discuss various aspects of diffeomorphism invariance of $O_g$, including the initial data at past timelike infinity. 

How and where we initialize the worldlines can be related to questions of diffeomorphism invariance, so we discuss our approach here. We imagine initializing on a surface $\Sigma$ that is taken to be arbitrarily far in the past $\tau_\Sigma \rightarrow -\infty$. This may be understood to model initial data determined by experiments in some laboratory. In the previous section, we solved
\begin{equation}
\frac{d^2}{d\tau^2}O_g^{(1)}(\tau) = - L\int_{\gamma(\alpha)} d\alpha \, R^{(1)}_{0n0n}(x^{(0)}(\tau,\alpha))
\end{equation} 
for $O_g^{(1)}$ using an $i\varepsilon$ prescription that determines boundary conditions of $O_g^{(1)}$ on $\Sigma$. This prescription is also relevant for integrating by parts to obtain the synchronous gauge expression. The boundary terms that arise vanish as a consequence of the $i\varepsilon$  prescription for which the associated Green's function is $G_R(\tau, \tau') = (\tau-\tau')e^{-\varepsilon(\tau-\tau')}\theta(\tau-\tau')$. In synchronous gauge, we have
\begin{align}
O_g^{(1)} =& -L\int_{-\infty}^\tau d\tau'(\tau-\tau')e^{-\varepsilon(\tau-\tau')} \int_{\gamma(\alpha)} d\alpha \, R^{(1)}_{0n0n} (x^{(0)}(\tau',\alpha)) 
\nonumber
\\
=& \frac{\kappa L}{2}\int_{\gamma(\alpha)} d\alpha \int_{-\infty}^\tau d\tau'(\tau-\tau')e^{-\varepsilon(\tau-\tau')}   \, \left(\frac{d}{d\tau'}\right)^2 h_{nn} (x^{(0)}(\tau',\alpha)),
\end{align}
where we used that $t$ and $\tau$ are equal up to $\mathcal{O}(h)$ corrections. Integrating by parts twice yields
\begin{equation}
O_g^{(1)}
= \frac{\kappa L}{2} \int_{\gamma(\alpha)} d\alpha \, h_{nn} (x^{(0)}(\tau,\alpha)),
\end{equation}
plus additional terms that are zero due to the exponential suppression of $e^{-\varepsilon(\tau-\tau')}$ as $\tau' \rightarrow -\infty$. Boundary terms arising from integrating by parts are therefore zero. Note that the $i\varepsilon$ here is precisely the $i \varepsilon$ introduced in \cite{Cheung:2026euf} to define the dressing there, and should similarly be understood here as part of the definition of our gravitational dressing. See \cite{Frob:2023awn, Goeller:2022rsx, Donnelly:2015hta, Giddings:2025xym} for further details of such setups. 

Now we can turn to studying diffeomorphism invariance of the expressions we derived for $O_g$, both in an arbitrary gauge and in synchronous gauge calculations. Classically, Eq. \eqref{OgArbitraryGauge} is already manifestly diffeomorphism invariant, as the linearized Riemann tensor is invariant. We can also check this explicitly within correlation functions by considering matrix elements with external gravitons, for example
\begin{equation}
\braket{0|O_g^{(1)}(\tau)|h(k)} = - L\int_{-\infty}^\tau d\tau' (\tau-\tau') \int_{\gamma(\alpha)} d\alpha \, \braket{0|R^{(1)}_{0n0n}(x^{(0)}(\tau',\alpha))|h(k)}.
\label{OgGravitonMatrixElementArbitraryGauge}
\end{equation}
This can be computed using $\braket{0|h_{\mu\nu}|h(k)} = e^{i k x} \epsilon_{\mu\nu}$ with polarization $\epsilon_{\mu\nu}$, where we suppress the polarization index. The familiar Ward identity applied to the external graviton still serves as a check on diffeomorphism invariance. It will however be useful to first make the equivalence between dressing and gauge fixing manifest \cite{Cheung:2026euf}. Presciently, we will work in momentum space, where acting on modes $\epsilon_{\mu\nu}e^{ikx}$ of the metric, the Riemann tensor can equivalently be written as 
\begin{align}
\frac{1}{(ik_0+\varepsilon)^2} R^{(1)}_{0n0n} 
&=
\frac{\kappa}{(i k_0+\varepsilon)^2}\left( - k_0k_n \epsilon_{0n}+\frac{1}{2}(k_n)^2 \epsilon_{00} + \frac{1}{2}(k_0)^2 \epsilon_{nn} \right)e^{ikx}  
\nonumber
\\
&= -\frac{\kappa}{2} P_{n\rho}P_{n \sigma}\epsilon^{\rho\sigma}e^{ikx},
\end{align}
where the projectors $P_{\mu \rho}$ are, in momentum space,
\begin{equation}
P_{\mu\rho} = \eta_{\mu\rho} - \frac{k_\mu \eta_{0\rho}}{k_0-i\varepsilon} = \eta_{\mu\rho}-\frac{k_\mu \dot{x}^{(0)}_{i,\rho}} {k\cdot \dot{x}^{(0)}_i -i\varepsilon},
\end{equation}
and we often leave the $\varepsilon$ implicit. The latter expression gives the projectors found in \cite{Cheung:2026euf} after accounting for the opposite choice of spacetime signature. In our case, however, these projectors arose simply as the representation of the Riemann tensor, which is gauge invariant, and not because the Riemann tensor is dressed.

Diffeomorphism invariance of the matrix element follows from invariance of the Riemann tensor matrix element. This can be checked the usual way,
by shifting $\epsilon_{\mu\nu} \rightarrow \epsilon_{\mu\nu} + \xi_\mu k_\nu + \xi_\nu k_\mu$,
\begin{align}
\braket{0|R^{(1)}_{0n0n}(x)|h(k)}  &\rightarrow \braket{0|R^{(1)}_{0n0n}(x)|h(k)} 
\nonumber
\\
&~~~~+\kappa
\left(-k_0k_n ( k_0 \xi_n + k_n \xi_0)+\frac{1}{2}(k_n)^2 (2 k_0\xi_0) + \frac{1}{2}(k_0)^2 (2 k_n \xi_n) \right) e^{ikx}
\nonumber
\\
&=\braket{0|R^{(1)}_{0n0n}(x)|h(k)}.
\end{align}
Invariance of the full matrix element, $\braket{0|O_g^{(1)}(\tau)|h(k)}$, is also manifest in the final expression,
\begin{align}
\braket{0|O_g^{(1)}(\tau)|h(k)} &= -L \int_{-\infty}^\tau d\tau' (\tau-\tau') \int_{\gamma(\alpha)} d\alpha \,
\braket{0|R^{(1)}_{0n0n}(x^{(0)}(\tau',\alpha))|h(k)}
\nonumber
\\
&=
\frac{\kappa L}{2}P_{n\rho}P_{n \sigma}\epsilon^{\rho\sigma} \int_{\gamma(\alpha)} d\alpha \,  e^{i k x^{(0)}(\tau, \alpha)},
\label{OgArbitraryGaugeMatrixElementProjector}
\end{align}
since $P_{\mu\rho}k^\rho =0$. The external graviton automatically obeys synchronous gauge, as $\dot{x}_i^{(0),\mu}P_{\mu\rho} =0$. 

One can also directly use synchronous gauge gravitons to compute the invariant matrix element. This aligns with the method to compute the dressed quantities via a particular gauge-fixing shown in \cite{Cheung:2026euf}, which in the present case amounts to using synchronous gauge for internal lines and the $P_{\mu\rho} P_{\nu \sigma}\epsilon^{\rho\sigma}$ for external polarizations. Indeed, this agrees precisely with Eq. \eqref{OgArbitraryGaugeMatrixElementProjector},
\begin{equation}
\braket{0|O_g^{(1)}|h(k)}
= \frac{\kappa L}{2} \int_{\gamma(\alpha)} d\alpha \, \braket{0|h_{nn} (x^{(0)}(\tau,\alpha)) |h(k)} = \frac{\kappa L}{2}P_{n\rho}P_{n \sigma}\epsilon^{\rho\sigma} \int_{\gamma(\alpha)} d\alpha \,  e^{i k x(\tau, \alpha)},
\end{equation}
as expected.

\section{Geometric observables in synchronous gauge}

To compute correlators of geometric observables, one must choose a gauge. Gauges in which the observer or apparatus plays a privileged role are especially transparent. One choice is synchronous gauge. This widely-used gauge makes comparison to the gravitational-wave literature straightforward. When computing correlators in quantum gravity, we will work in transverse-traceless (TT) gauge, which is a special case of synchronous gauge. Here we compute the geometric observables and their correlators in synchronous gauge.

We describe the setup here. We expand about flat space in the synchronous gauge,
\begin{align}
\label{metric_synchronous_gauge}
ds^{2}= -dt^{2}+ (\delta_{ij}+\kappa h_{ij})dx^{i}dx^{j},
\end{align}
and work to first order in $\kappa h$. Explicitly, the synchronous gauge conditions on $h_{\mu\nu}$ are
\begin{equation}
\label{syschronous_gauge}
h_{00}=h_{0i}=0.
\end{equation}
In synchronous gauge, it is known that timelike geodesics with zero initial velocity, like the worldlines $x_1, x_2$ as depicted in Fig. \ref{fig:geodesics}, remain in the same spatial location to all orders in $h$ for perturbations around Minkowski. 

\label{app:geometric_observables_synchronous}

\subsection{Proper distance}
\label{app:proper_distance_synchronous}

In this Appendix, we compute the spacelike distance $\sigma$ between two geodesics and then the leading graviton correction to the vacuum two-point function of $\sigma$. 

We begin by defining proper distance $\sigma(\tau_1,\tau_2) = \sigma(x_1(\tau_1),x_2(\tau_2))$ classically
as the geodesic distance between points $x_1(\tau_1),x_2(\tau_2)$.\footnote{In the main text, we set $\tau_1 = \tau_2$ for simplicity.} This definition of $\sigma$ holds perturbatively to leading order even for an off-shell metric perturbation, which is the order to which we work. 

We can parametrize the proper times of worldlines $x_1$ and $x_2$ via coordinate times $t_1, t_2$. We also consider a spacelike geodesic $x(\lambda)$ that connects $x_1(t_1)$ and $x_2(t_2)$, whose proper length is $\sigma(t_1,t_2)$. This geodesic is the wavy curve in Fig. \ref{fig:geodesics}. We can expand perturbatively in powers of $h$ as
\begin{equation}
\label{spacelike_geodesic_perturb_exp}
x(\lambda)= x^{(0)}(\lambda)+ x^{(1)}(\lambda)+ x^{(2)}(\lambda)+\dots,
\end{equation}
where the affine parameter $\lambda$ parametrizes the spacelike geodesic with boundary conditions
\begin{align}
\label{pert_boundary_conditions_spacelike_curve}
x^{(0)}(0)&= (t_1,\mb x_1)~,~x^{(0)}(1)= (t_2,\mb x_2),\nonumber\\
x^{(1)}(0)&=0~~,~~x^{(1)}(1)=0.
\end{align}
We use bolded notation to denote spatial vectors. 

We can then perturbatively solve the geodesic equation
\begin{equation}
\label{geodesic_equation}
\ddot{x}^\mu +\Gamma\indices{^{\mu}_{\rho \sigma}}\dot{x}^\rho \dot{x}^\sigma =0
\end{equation}
to determine $x$, which is depicted as the wavy curve in Fig. \ref{fig:geodesics}. At leading order, the geodesic equation \eqref{geodesic_equation} gives the straight line solution $\ddot x^{(0)}=0$,
\begin{equation}
\label{zeroth_order_spacelike_geod_solution}
x^{(0),0}(\lambda)= t_1+ \lambda (t_2- t_1)~,~\mb x^{(0)}(\lambda)= \mb x_1+\lambda (\mb x_2- \mb x_1).
\end{equation}
This zeroth order solution is all we need to define the leading correction to $\sigma$. We perturbatively expand the geodesic distance
\begin{equation}
\label{spacelike_geodesic_distance_lambda}
\sigma(t_1, t_2)= \int_{0}^{1} d\lambda \sqrt{g_{\mu\nu}\dot x^{\mu}\dot x^{\nu}}, 
\end{equation}
which to leading order gives 
\begin{equation}
\label{sigma_leading_order_expression}
\sigma(t_1, t_2)=L_\tau  + \frac{\kappa}{2L_\tau}  \int_{0}^{1}d\lambda \, h_{ij}(x^{(0)}(\lambda))\dot x^{(0),i}\dot x^{(0),j},
\end{equation}
where as a reminder, $L_\tau=\sqrt{(\mb x_2 - \mb x_{1})^2 - (t_2-t_1)^2}$.

Now that we defined $\sigma$ classically, we turn to computing its vacuum two-point function where $h$ is quantized according to the rules of quantum field theory. We will be interested in computing Wightman functions of $\sigma$. In this case, it is equally straightforward to use either the canonical quantization approach in \cite{Carney:2024wnp} and work in TT gauge directly, or equivalently the path integral approach in \cite{Cheung:2026euf, Sivaramakrishnan:2025srr}, which in this special case amounts to working in synchronous gauge while $h$ is off-shell and then using the residual gauge freedom present for Wightman (on-shell) correlators to then specialize to TT gauge for the Wightman function. We will take the latter approach to provide more general expressions for future use, although for the correlators we evaluate here, the differences will be irrelevant. This procedure was also carried out in \cite{Sivaramakrishnan:2025srr, Cheung:2026euf} for other observables.

The two-point function of $\sigma$ can be defined via the path integral. Using $\hbar=c=1$ units,
\begin{equation}
\braket{\sigma(\tau_1,\tau_2)\sigma(\tau_3,\tau_4)} = \int \mathcal{D}[h] \sigma(\tau_1,\tau_2)\sigma(\tau_3,\tau_4) e^{i S_{\rm EH}[h]},
\end{equation}
together with the appropriate boundary conditions or $i\varepsilon$ prescription that specifies the type of correlator.

To calculate the connected vacuum correlator $\braket{\sigma(t-i\varepsilon,t-i\varepsilon)\sigma(0,0)}_c$ we use the TT gauge graviton mode expansion
\begin{equation}
\label{graviton_mode_expansion_main}
h_{\mu\nu}(x) =\sum_{s} \int \frac{d^3\mb{k}}{ \sqrt{(2\pi)^3 2 E_{\mb{k}}}} \left[ \varepsilon^{s}_{\mu\nu}(\mb{k}) e^{i k \cdot x} a_{\mb{k},s} + {\rm h.c.} \right],
\end{equation}
where $s=1,2$ labels polarization states, and $k\cdot x= - E_{\mb k}t + \mb k\cdot \mb x$, with $E_{\mb k}=|\mb k|$. The canonical commutation relations are  $[a_{s,\mb k},a^{\dagger}_{s',\mb k'}]=\delta_{s,s'}\delta(\mb k - \mb k')$. After performing the $k$-integral of Eq.~\eqref{sigma_sigma_correlator_PSD_form}, and setting $\tau_1=\tau_2$ for simplicity, we obtain 
\begin{align}
\label{connected_spacelike_two_point_correlator_result}
&\langle \sigma (t- i\varepsilon)\sigma (0)\rangle_{c}
\nn
\\
&~~~=\frac{\kappa^{2}}{72\pi^2 }\bigg[17 - \frac{3 t^2}{L^2}- \frac{3}{L^3}\left(\frac{3L^4 + 6 L^2 t^2 - t^4}{(t-i\varepsilon)}\right)\text{ArcTanh}\left(\frac{L}{t- i\varepsilon}\right) 
- 12 \log\left(1+ \frac{L^2}{(it + \varepsilon)^{2}}\right)\bigg]~ ,
\end{align}
with $\varepsilon \to 0^{+}$. Taking into account the $\varepsilon \to 0^{+}$ branch cuts we can arrive at a compact expression of the correlator in terms of $s=t/L$ for any $t$
\begin{align}
\label{connected_spacelike_two_point_correlator_real_imaginary_result}
\lim_{\varepsilon \to 0^{+}}\langle \sigma (s-i\varepsilon)\sigma (0)\rangle_{c}=&\frac{\kappa^{2}}{72\pi^2 }\bigg(17 - 3s^2 + \frac{9\pi^2 }{2}\delta(s) -12 {\cal P}\left(\log \big| 1- \frac{1}{s^2} \big|\right)  \nonumber\\&
- \frac{3}{2}(3 + 6 s^2 - s^4){\cal P}\left(\frac{1}{s}\log \big| \frac{s+1}{s-1} \big|\right)\bigg)\nonumber\\&
+ i\frac{\kappa^{2}}{72\pi }\bigg(- \frac{3}{2}(3 + 6 s^2 - s^4){\cal P}\left(\frac{1}{s}\Theta(1-|s|)\right) \nonumber\\&
+ 12\, \text{sgn}(s )\,\Theta\left(1- |s|\right)\bigg),
\end{align}
where $\Theta (x)$ is the Heaviside step function and $\cP(\cdot)$ represents the Cauchy principal value.
In Section \ref{subsec:proper_distance}, we presented the above formula for the special cases when $t>L$ and $0<t<L$.

\subsection{Arrival time}
\label{app:arrival_time_synchronous}

We will derive the arrival time quantity in synchronous gauge using the metric Eq. \eqref{metric_synchronous_gauge}. Let us define the light cone coordinate 
\begin{equation}
u \equiv t-z.
\end{equation}
We consider a light-ray $x(\eta)$, where $\eta \in [\eta_i, \eta_i+L]$ is an affine parameter. Let us perturbatively expand the light-ray in powers of $h$,
\begin{equation}
x(\eta)= x^{(0)}(\eta)+ x^{(1)}(\eta)+\dots,
\end{equation}
with zeroth order data
\begin{equation}
\label{initial conditions_light_ray}
u^{(0)}(\eta_{i})=u_{0},\quad \dot x^{\mu,(0)}(\eta_{i})=(1,0,0,1).
\end{equation}
The second condition above implies $\dot u^{(0)}(\eta_{i})=0$. 
The geodesic equation for the light ray $\ddot x^{\mu}+ \Gamma\indices{^{\mu}_{\nu\rho}}\dot x^{\nu}\dot x^{\rho}=0$ implies $\ddot x^{(0)}(\eta)=0$ and using the boundary conditions above we obtain without loss of generality
\begin{equation}
x^{(0)}(\eta;u_{0})= (\eta,0,0,\eta-u_{0} ).
\label{unperturbed_light_path}
\end{equation}
We see that $u^{(0)}=u_{0}$ for the unperturbed light ray. From the null condition $ds^2=0$ we obtain to $O(h)$
\be
dz= dt \left(1-\frac{\kappa}{2}h_{zz}\right).
\ee
Integrating the above along the light-ray we find
\be
t-z= t_i- z_{i}+ \frac{\kappa}{2}\int_{t_{i}}^{t}dt'~ h_{zz}(x(t')).
\ee
To leading order we can set $t= t^{(0)}(\eta)$ in the upper limit of the integral and use the unperturbed path $x^{(0)}(t;u_{0})$ in the integrand. \eqref{unperturbed_light_path} allows us to change integration variable from $t'$ to $\eta'$, and using the definitions $u(\eta)=t(\eta)-z(\eta)$ and $u_{0}= t_i-z_i$, we obtain Eq. \eqref{vshift},
\begin{equation}
u(\eta)= u_{0}+\frac{\kappa}{2}\int_{\eta_{i}}^{\eta}d\eta' h_{zz} (x^{(0)}(\eta';u_{0}))+ O(h^2). 
\end{equation}
The explicit expression for the vacuum two-point function can be found in \cite{Sivaramakrishnan:2025srr}.

\subsection{Round trip time}
\label{app:round-trip_time_synchronous_gauge}

The classical computation of round trip time is exceedingly well-studied, so here we simply give an overview of various results in the literature. 

In the study of gravitational waves, the round trip time is well studied and usually computed in transverse-traceless (TT) gauge. For example, see \cite{Maggiore:2007ulw}. This gauge applies to on-shell configurations, like gravitational waves. This gauge is especially convenient for computations of correlation functions, as the TT gauge propagators take a simple form. 

When we perform the path integral to compute quantum gravity corrections, it is essential that the operator is defined for off-shell $h$. Synchronous gauge is therefore used. For on-shell $h$, synchronous gauge possesses a further gauge freedom, which can be fixed to yield TT gauge. The synchronous gauge computation is a straightforward exercise, for example see \cite{Sivaramakrishnan:2025srr}. See also \cite{Lee:2024oxo}, in which the classical round trip time was computed in an arbitrary gauge in terms of $h$ and then reduced to synchronous gauge as a special case. The vacuum Wightman two-point function has on-shell $h$, and so can be computed using the TT gauge graviton two-point function. 

The vacuum two-point function of proper time was computed and studied in \cite{Bonga:2013uha,Carney:2024wnp ,Sivaramakrishnan:2025srr, Freidel:2026hed} from various points of view, including the path integral approach we employ here. We direct the reader to these works for details and exact expressions. 

\newpage

\bibliography{references}{}
\bibliographystyle{utphys}

\end{document}